\documentclass{aa}
\usepackage{graphicx}
\usepackage{txfonts}
\usepackage{lipsum}
\usepackage{subfiles}
\usepackage{subcaption}         % necessary for continued figures, example in section 3
\usepackage{lscape}             % to rotate a single page table, example in appendix.
\usepackage{placeins}           % useful with \FloatBarrier, to keep 
\usepackage{siunitx}   
\usepackage{hyperref}
\begin{document}

   \title{Generalizing Across Astronomical Surveys: Few-Shot Light Curve Classification with Astromer 2}

   \author{Cristobal Donoso-Oliva\inst{1, 3, 5}
        \and Ignacio Becker\inst{2}
        \and Pavlos Protopapas\inst{2}
        \and Guillermo Cabrera-Vives\inst{1,3,4,5,6}\fnmsep
        \and Martina Cádiz-Leyton\inst{1, 3}
        \and Daniel Moreno-Cartagena\inst{1, 3}
        }

   \institute{
Department of Computer Science, Universidad de Concepción, Edmundo Larenas 219, Concepción, Chile\and
John A. Paulson School of Engineering and Applied Science, Harvard University, Cambridge, MA, 02138\and
Center for Data and Artificial Intelligence, Universidad de Concepción, Edmundo Larenas 310, Concepción, Chile\and
Millennium Institute of Astrophysics (MAS), Nuncio Monseñor Sotero Sanz 100, Of. 104, Providencia, Santiago, Chile\and
Millennium Nucleus on Young Exoplanets and their Moons (YEMS), Chile\and
Heidelberg Institute for Theoretical Studies, Heidelberg, Baden-Württemberg, Germany
   }

   \date{Received Novemember xx, 2024}

% \abstract{}{}{}{}{}
% 5 {} token are mandatory
 
  \abstract{
  %context heading (optional)
   Foundational models have emerged as a powerful paradigm within the deep learning field. Their capacity relies on the ability to learn robust representations from large-scale datasets and generalize to diverse downstream applications, such as classification. In this paper, we present Astromer 2, a foundational model designed for extracting light curve embeddings.}
   {We introduce Astromer 2, an enhanced iteration of our self-supervised model for light curve analysis. This paper highlights the advantages of its pre-trained embeddings, compares its performance with that of its predecessor, Astromer 1, and provides a detailed empirical analysis of its capabilities, offering deeper insights into the model’s representations.} 
   {Astromer 2 is pretrained on 1.5 million single-band light curves from the MACHO survey using a self-supervised learning task that predicts randomly masked observations within sequences. Fine-tuning on a smaller labeled dataset allows us to assess its performance in classification tasks. The quality of the embeddings is measured by the F1 score of an MLP classifier trained on Astromer-generated embeddings.}
   {Our results demonstrate that Astromer 2 significantly outperforms Astromer 1 across all evaluated scenarios, including limited datasets of 20, 100, and 500 samples per class. The use of weighted per-sample embeddings, which integrate intermediate representations from Astromer’s attention blocks, is particularly impactful. Notably, Astromer 2 achieves a 15\% improvement in F1 score on the ATLAS dataset compared to prior models, showcasing robust generalization to new datasets. This enhanced performance, especially with minimal labeled data, underscores the potential of Astromer 2 for more efficient and scalable light curve analysis.}
    {}

   \keywords{Representation Learning --
             Light Curves --
             Foundational Models
               }

   \maketitle

%%%%%%%%%%%%%%%%%%%%%%%%%%%%%%%%%%%%%%%%%%%%%%%%%%%%%%%%%%%%%%
\section{Introduction}
%Light curve analysis has been a cornerstone in astronomy for characterizing stellar objects \citep{deb2009light}.
Light curve analysis is a cornerstone in astronomy for characterizing stellar objects \citep{deb2009light}. By analyzing the time-series data of luminosity variations, astronomers can extract statistical features that enable classification and identification tasks \citep{richards2011machine}.

Although traditional methods show success \citep{sanchez2021alert, chaini2024light}, the advent of foundational models presents fresh opportunities to gain insights into cosmic variability. Foundational models are deep neural networks trained using self-supervised techniques on extensive datasets \citep{bommasani2021opportunities, awais2023foundational}. These models acquire a thorough grasp of their domain, allowing for the creation of versatile representations applicable to various downstream tasks.

We note that self-supervised learning by itself does not necessarily define a \textit{foundation model}. Foundation models typically involve large-scale pretraining on massive datasets combined with the ability to generalize and adapt effectively across a wide range of downstream tasks, often through fine-tuning. Our approach aligns with this broader concept, but self-supervision is one component within that larger framework

Self-supervised learning enables the use of vast amounts of unlabeled data, significantly increasing the effective training volume beyond what is possible with labeled samples alone. This expanded training capacity allows models to learn richer, more generalizable representations that improve downstream performance even when labeled data is scarce.

Classical techniques rely on manually engineered features \citep{debosscher2007automated, nun2015fats}, which may introduce biases or fail to capture intricate patterns \citep{2022MNRAS.517.3660P}. Foundational models, by processing large volumes of data, have the potential to reveal novel, precise structures in the data. However, this gain in representational power comes at the expense of reduced interpretability. While self-supervised deep learning models can offer powerful representations, their interpretability remains limited compared to mechanistic, physics-based models that have clear, physically grounded parameters. However, their interpretability is comparable to other deep learning approaches.

To contextualize our work within the broader field, we reference recent advances in time-series self-supervised learning, such as the Foundation Models for Time Series \citep{liang2024foundation}, which explore similar concepts in the machine learning community.

In 2023, we introduced Astromer, a self-supervised model designed to extract general-purpose embeddings from light curves 
\citep{astromer}. Trained on 1.5 million light curves, Astromer demonstrated consistent improvements in classification tasks compared to models trained directly on labeled datasets.

The motivation for using self-supervised learning in Astromer is consistent with that of foundational models across AI, including large language models and some of the vision transformers. In the context of astronomical time-series, labeled data are often scarce and costly to obtain at scale. Self-supervised masked modelling enables leveraging vast amounts of unlabeled data to learn powerful, general representations that can be fine-tuned with limited labels for downstream tasks such as classification. This approach has become a central paradigm in modern AI and guides the design of Astromer too.

Our self-supervised masked-modelling approach is inspired by successful techniques in natural language processing (e.g., BERT) and computer vision (e.g., ViTs), which have demonstrated the power of learning rich representations by reconstructing missing parts of the input.

Other foundational models in astronomy, such as those employing contrastive learning \citep{lanusse2023astroclip, rizhko2024self, parker2024astroclip}, integrate multiple data modalities to create richer and more complex representations. While multi-modal learning is a promising avenue, it introduces additional complexity in model training and interpretation \citep{wang2024comprehensive}.

Astromer, by contrast, focuses solely on single-modality light curve data, leveraging its temporal structure without requiring alignment or integration steps across modalities. Instead of contrastive learning, Astromer employs magnitude imputation to handle missing values in time series, resulting in a simpler, yet highly effective model that achieves state-of-the-art performance without incurring high computational costs.

In this paper, we present Astromer 2, an improved version of our original model. For consistency, we use the same dataset from our initial publication and compare our latest model in classification task. Additionally, we delve into the embedding vectors and attention weights to better understand the model’s capabilities and performance as used in other works \citep{10.1093/mnras/stab2588}.

In this work, Section \ref{sec:astromer0} revisits the main characteristics of the original version of Astromer, namely Astromer 1. While a detailed explanation is available in our previous paper, we dedicate a significant portion of this section to reiterating its architecture for several key reasons. Firstly, much of the formulation of Astromer 2 is fundamentally built upon the principles of Astromer 1. Secondly, in response to community questions and to reflect our own improved understanding of its internal processes, we have rewritten the description of the architecture in a clearer and more concise manner. With this enhanced explanation, we aim for this paper to be self-contained, allowing readers to fully comprehend our work without referencing the previous publication. Section \ref{sec:astromer1} then introduces the improvements that define the new model. Sections \ref{sec:data} and \ref{sec:results} describe the data and the main results of this work, respectively. Finally, Section 6 outlines the main findings and conclusions.
%In this work, Sect. \ref{sec:astromer0} describes the main characteristics of the original version of Astromer, namely Astromer 1. Section \ref{sec:astromer1} introduces the improvements to the model. Sections \ref{sec:data} and \ref{sec:results} describe the data and the main results of this work, respectively. Finally, Sect. \ref{sec:conclusion} outlines the main findings and conclusions.
\section{Astromer 1}\label{sec:astromer0}
The initial version of Astromer \citep{astromer} adapted the BERT text model from natural language processing \citep{devlin2018bert}. While both light curves and text are sequential data, light curves pose a unique challenge due to the inherent irregularities in their sampling. Moreoever,  instead of a discrete vocabulary, we work with continuous magnitudes for each token. 

Despite the differences between BERT and Astromer, the high-level approach to training remains similar, as it leverages a self-supervised task. 
Specifically, we employ a masking strategy that obscures portions of the light curve, allowing the model to predict the missing magnitudes. This technique, inspired by BERT’s word masking in sentences, enables the model to learn meaningful representations without relying on human-annotated labels.

This section revisits the pipeline previously introduced in Astromer 1. While much of the content has been explained before, we present it here with a more refined and clearer explanation for enhanced understanding.

% =============================================================
% DATA PREPARATION ============================================
% =============================================================
\subsection{Data preparation}\label{sec:data_preparation}
Astromer uses single-band light curves $\{x_i\}_{i=0}^{N-1}$ with $N$ as the number of samples. Each sample is represented as a set of tuples $x_i = \{(t_l, m_l, e_l)\}_{l=0}^{L_i-1}$. Here, $t_l$ denotes the observation time in modified Julian date (MJD), $m_l$ represents the magnitude, and $e_l$ corresponds to the magnitude uncertainty. The maximum number of observations $L_i$ varies across samples, resulting in a variable-length dataset. 

We fixed a maximum length of 200 observations to create the network’s input. During pretraining, we sample different windows of 200 observations per epoch, allowing the model to see most of the light curve sequence in small, fixed chunks. Shorter light curves are zero-padded to a fixed length of 200 observations.

After constructing the windows, we normalize their values. Specifically, we subtract $\bar{x_i} = (\bar{t}, \bar{m}, \bar{e})$ the mean value of each light curve, producing zero-mean samples with non-scaled amplitude. Our experiments have shown that this normalization step is essential for the model to converge effectively. Other options may be insufficient to produce valuable embeddings.

% =============================================================
% INPUT EMBEDDING  ============================================
% =============================================================
\subsection{Input Embedding}\label{sec:input_embedding}
Unlike language models, Astromer does not have a fixed vocabulary of tokens. Instead, the input consists of a sequence of continuous magnitude values, each paired with its corresponding observation time in MJD. We do not consider the uncertainties in Astromer's input.

To create a single input embedding, we transform each time and magnitude scalar into vectors. To encode observation times, we apply an adapted positional encoder (PE) that scales the angular frequencies $\omega_j$ using the observation time $t_l$, capturing the irregular sampling in the temporal representation.
\begin{equation}\label{eq:pe}
\text{PE}_{\textit{j}, t_{l}} = \left\{
        \begin{array}{ll}
            \sin{(t_{l}\cdot\omega_{\textit{j}})} & \quad j \text{ is even} \\
            \cos{(t_{l}\cdot\omega_{\textit{j}})} & \quad j \text{ is odd}
        \end{array}
    \right.
\end{equation}
In Eq. \ref{eq:pe}, $\textit{j} \in [0,...,d_{pe}-1] $, where $d_{pe}=256$ is the PE dimensionality and $\omega_\textit{j}$ is the angular frequency defined as,
\begin{equation}\label{eq:angular_freq}
    \omega_{\textit{j}} = \frac{1}{1000^{2\textit{j}/d_{pe}}}.
\end{equation}

For the magnitudes, we linearly project each magnitude value into a vector of size $d_{pe}=256$. The weights used to transform magnitudes are initialized from a Normal distribution and subsequently learned during training. This shared set of weights is applied across all observations.

The final input embedding $\rm{X}\in\mathbb{R}^{L\times 256}$ is the sum of the PE and transformed magnitudes, $\rm{X} = PE + \boldsymbol{m}\rm{W}^{\top}$, where $\boldsymbol{m} \in \mathbb{R}^{L\times 1}$ is the magnitudes vector and $\rm{W} \in \mathbb{R}^{256\times 1}$ are the weights for transforming scalars into vectors.

% =============================================================
% PROBED AND MASKING ==========================================
% =============================================================
\subsection{Probed and Masking}
The key to Astromer learning a good representation is to pretrain it to predict unseen observations along the light curve sequence. The probed subset consists of magnitudes designated for the model to predict. However, these values are excluded when calculating attention weights. We randomly select $50\%$ of the total observations per window to constitute the probed subset. This subset is denoted by a binary mask vector, where the 1's correspond to the probed magnitudes, and zero otherwise.

\begin{figure}[!ht]
    \centering
    \includegraphics[scale=0.30]{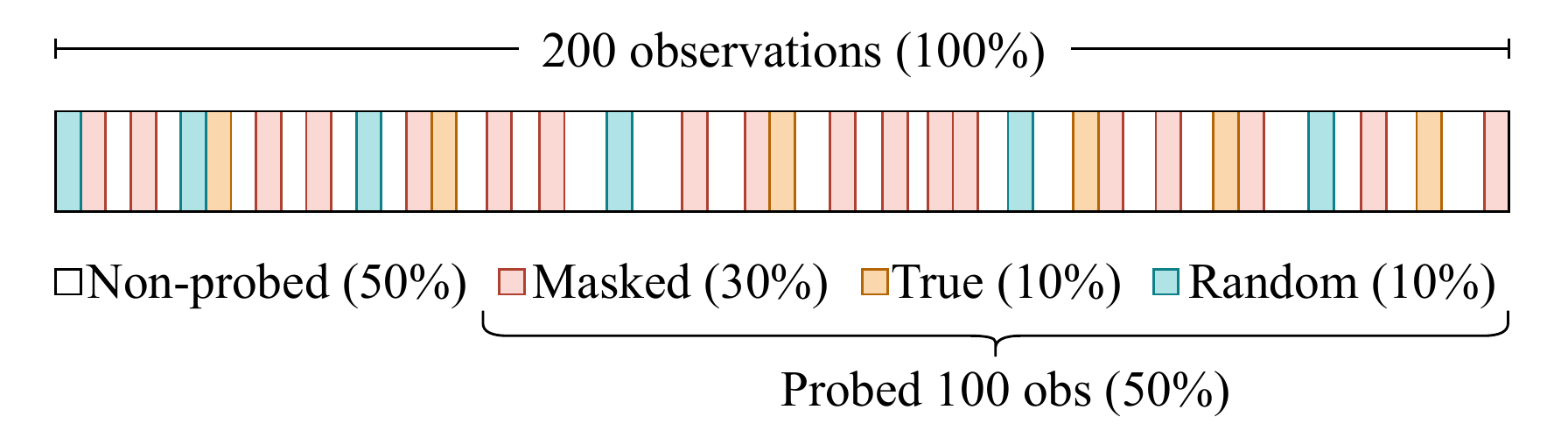}
     \caption{The self-supervised masking strategy used for pretraining. For each light curve, 50\% of the observation points are selected as the 'probed' subset, which the model must predict. This subset consists of three components: 30\% of the points are fully masked (hidden), 10\% are replaced with random magnitudes, and 10\% remain visible. This strategy forces the model to learn from context rather than simply memorizing positions.}
    % \caption{Observation subsets. In our pretraining strategy, $50\%$ of the observations are designated as probed. The probed subset is evaluated in the loss function to optimize Astromer's parameters. Its composition emphasizes predicting hidden elements (with $30\%$ masked overall) while mitigating the risk of the model over-relying on the masked observations ($10\%$ visible and $10\%$ with random magnitudes). This approach ensures the model pays attention to current observations without overfitting to them.}
    \label{fig:probed_masking}
\end{figure}

In the self-attention mechanism, the attention weights for the probed subset are set to zero using masking. This design encourages the model to leverage the surrounding context to predict the probed magnitudes. During inference, however, masking is not applied. To prevent the model from over-relying on the masked observations during training, we adopt a the following strategy. We assign $10\%$ of visible observations and $10\%$ of random observations in the probed subset. As a result, the actual masked portion is reduced to $30\%$, while the probed subset still corresponds to the initial $50\%$. This approach mitigates the risk of the model learning a direct identity mapping and improves its robustness to noise. Figure \ref{fig:probed_masking} illustrates the composition of the final subsets,  with the probed subset size fixed at $50\%$ of the observations.

% =====================================================
% ENCODER  ============================================
% =====================================================
\subsection{Encoder}
\begin{figure}[!ht]
    \centering
    \includegraphics[scale=0.45]{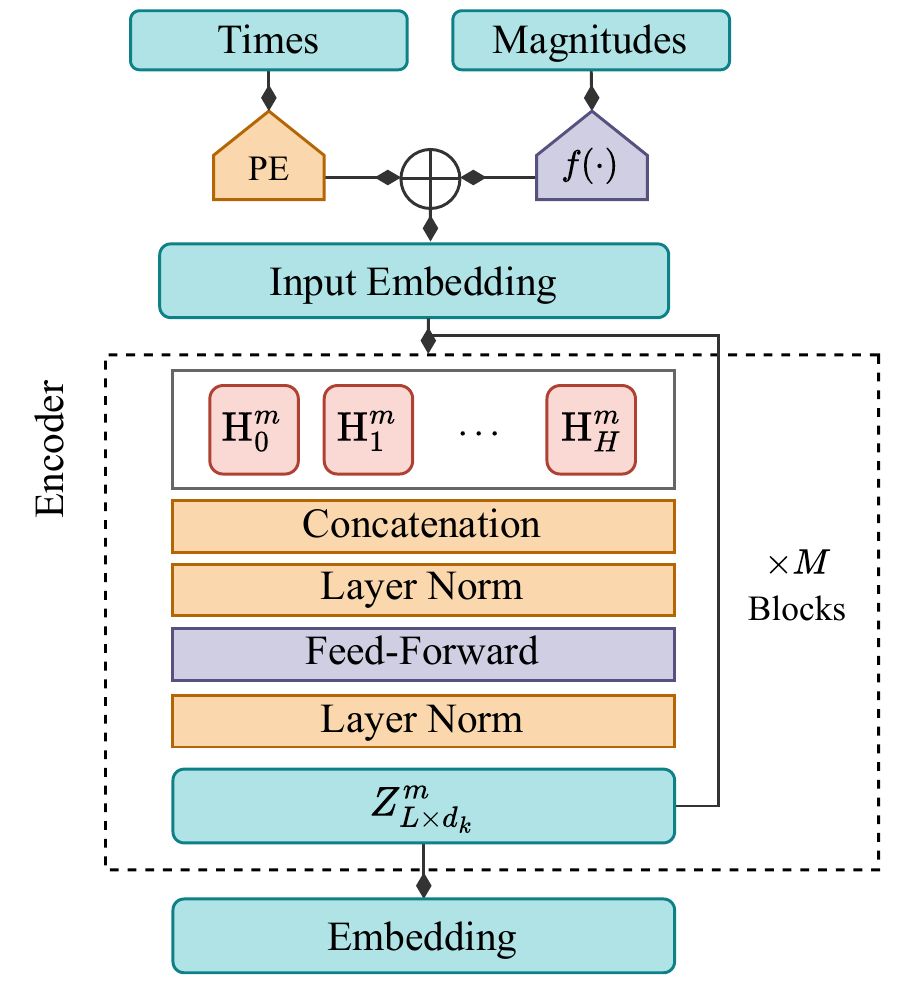}
    \caption{Overview of the Astromer 1 architecture. An input embedding is formed by summing a positional encoding (PE) of the observation times and a linear projection of the magnitudes. This embedding is processed by an encoder composed of M=2 blocks, each containing H=4 self-attention heads, to produce the final light curve representation, which is derived from the output of the last block.}
    % \caption{Astromer 1 architecture diagram. The input embedding is the sum of PEs for times and linearly transformed magnitudes. In Astromer 1, the encoder comprises $M=2$ blocks with $H=4$ attention heads, each having $64$ units. The outputs of the attention heads are concatenated and processed through a feed-forward network. The final embedding is derived from the output of the last attention block.}
    \label{fig:astromer_0}
\end{figure}

The encoder comprises a sequence of attention blocks connected in series. The first block processes the input embeddings described in Sect. \ref{sec:input_embedding} and a binary mask matrix that specifies which observations to exclude from the attention mechanism. Subsequent blocks take as input the output of the preceding attention block as shown in Fig. \ref{fig:astromer_0}.

Astromer 1 has two attention blocks, each containing four heads with 64 units. The outputs of the heads are concatenated, normalized, and combined through a fully connected layer, one hidden layer of 128 units, and a hyperbolic tangent activation. 

Within each head, the attention values are computed from the similarity matrix derived using the Query (Q), Key (K), and Value (V) matrices, normalized by the square root of $d_k=256$, the model’s embedding size. Note that Q, K, and V come from a linear transformation of the input embedding and have different values for each head and block.
\begin{eqnarray}\label{eq:att_weights}
    \label{eq:mask_selfatt}
    \rm{Q}&=& \rm{X}\rm{W}_{query}^{\top}\hspace{6mm} \nonumber
    \rm{K} \hspace{2mm}=\hspace{2mm} 
    XW_{key}^{\top}\hspace{6mm}\rm{V} \hspace{2mm}=\hspace{2mm} \rm{X}\rm{W}_{value}^{\top}\\
    \rm{W}_{att} &=& \text{Softmax} \left ( \frac{\rm{Q}\rm{K}^{\top} -\infty\text{M}}{\sqrt{d_{k}}}\right )\\
    \rm{Z} &=& \rm{W}_{att}\rm{V}\nonumber
\end{eqnarray}
In Eq. \ref{eq:att_weights}, the mask matrix $\rm{M}$ prevents the masked subset of probed observations from contributing to the attention values. When $\rm{M}=1$, the argument of the softmax function is effectively negative infinity, resulting in a zero attention weight.

In Astromer 1, the output of the last block serves as the final embedding. This matrix has two functions: reconstructing magnitudes during pretraining and serving as the embedding for downstream tasks.

% =====================================================
% PRETRAINING TASK ====================================
% ===================================================== 
\subsection{Pretraining Task}
We pretrain the model to predict the magnitudes of the probed subset in each input sequence. This is achieved by passing %$\boldsymbol{z}\in\mathbb{R}^{200\times256}$, 
the output embedding from the last attention block, through a fully connected network with no hidden layers or activation. The result is a vector of estimated magnitudes, $\boldsymbol{\hat{x}} \in \mathbb{R}^{200\times 1}$, providing the reconstruction for each time point according to its related embedding.

We constraint the loss function to compute the root-mean-square error on the probed subset only:
\begin{eqnarray}\label{eq:loss}
    \mathcal{L}oss = \sqrt{\frac{1}{N-1}\sum_{i=0}^{N-1}\sum_{l=0}^{L-1} m_{il}(x_{il} - \hat{x}_{il})^2}.
\end{eqnarray}
In Eq. \ref{eq:loss}, $N$ represents the number of training samples, and $L=200$ represents the length of the windows. Thus, the masking vector $\boldsymbol{m}_{i}$ selectively includes errors from the probed subset.

\section{Astromer 2}\label{sec:astromer1}
Astromer 2 incorporates features that were not included in the initial version due to resource and time constraints. While Astromer 1 served as a proof of concept for generating effective embeddings, Astromer 2 builds on this foundation, introducing iterative enhancements to optimize performance at each stage.

\begin{figure}
    \centering
    \includegraphics[scale=0.45]{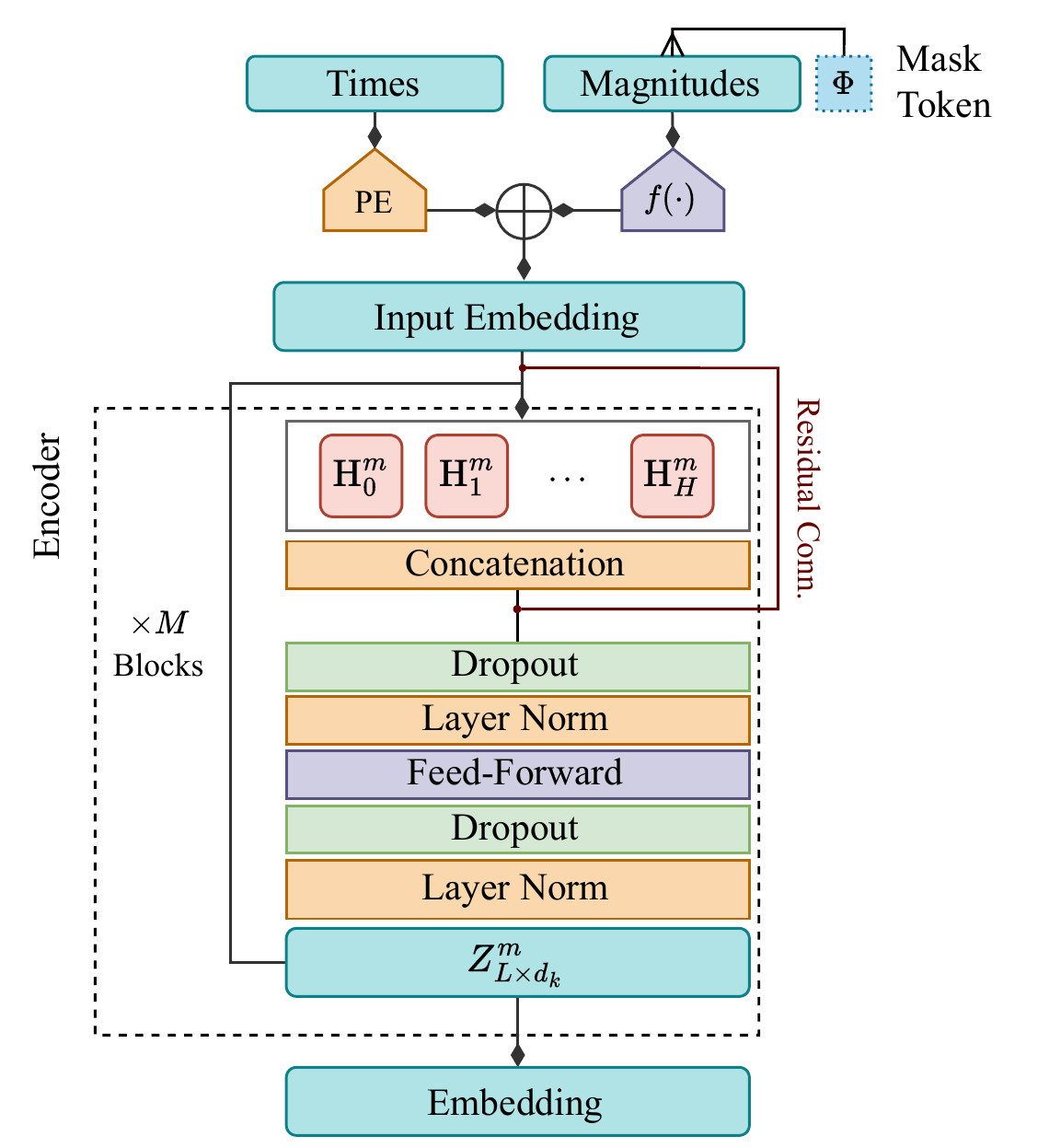}
     \caption{The Astromer 2 architecture, an enhanced version of the model shown in Fig. \ref{fig:astromer_0}. The primary architectural change is at the input stage: magnitudes designated for masking are now replaced by a single, trainable MASK token. Additionally, the encoder's depth is increased to M=6 blocks to improve its representational capacity.}
    \label{fig:astromer_1}
\end{figure}

Building upon the foundation of Astromer 1 discussed in Sect. \ref{sec:astromer0}, this section is dedicated solely to the enhancements that distinguish the principal features of Astromer 2. Fig. \ref{fig:astromer_1} shows a visual representation of the updated architecture of Astromer 2.

% ====================================================================================
% INPUT EMBEDDING ====================================================================
% ====================================================================================
\subsection{Input embedding}
The process for creating the input embedding for Astromer 2 remains the same as in the initial version. However, we replace the magnitudes targeted for masking with a trainable token that is zero-initialized and shared across all samples.

While the contribution of masked tokens is zero after the attention weight calculation, adding a mask token to replace the actual magnitude allows the model to recognize which tokens are masked, which can be helpful during training. We also avoid potential information leaks that could arise from the all-to-all computation within the similarity matrix.

% ====================================================================================
% ENCODER  ===========================================================================
% ====================================================================================
\subsection{Encoder}
The encoder of Astromer 2 has a significantly larger number of parameters, increasing from \num{661505} to \num{3953409}. A six fold increase. This growth is due to the inclusion of six attention blocks, with each block containing four heads and 64 units. Additionally, we have incorporated a dropout layer after the self-attention calculation, as depicted in Fig. \ref{fig:astromer_1}.

% ====================================================================================
% PRETRAINING TASK ===================================================================
% ====================================================================================
\subsection{Pretraining task}
Like Astromer 1, we use the root-mean-square error as the loss function. In Astromer 2, however, the losses are scaled based on observational uncertainties. These uncertainties are normalized to a range of 0 to 1, and their reciprocals are used as weights. Incorporating this scaling term into the error calculation enhances performance compared to Astromer 1.
\begin{eqnarray}\label{eq:loss_a1}
    \mathcal{L}oss = \sqrt{\frac{1}{N-1}\sum_{i=0}^{N-1}\sum_{l=0}^{L-1} \frac{m_{il}}{e_{il}}(x_{il} - \hat{x}_{il})^2}.
\end{eqnarray}
In Eq. \ref{eq:loss_a1}, $e_{il} \neq 0$ denotes the observation uncertainty associated with step $l$ in window $i$.
\section{Data Sources}\label{sec:data}
In this section, we introduce our training data, including unlabeled light curves for pretraining and labeled samples for the downstream classification task. 

\subsection{Unlabeled data - MACHO}
The MACHO project \citep{1993Natur.365..621A} aimed to detect Massive Compact Halo Objects (MACHO) to find evidence of dark matter in the Milky Way halo by searching for gravitational microlensing events. Light curves were collected from 1992 to 1999, producing light curves of more than a thousand observations \citep{1999PASP..111.1539A} in bands B and R.
The observed sky was subdivided into 403 fields. Each field was constructed by observing a region of the sky or tile. The resulting data is available in a public repository\footnote{\url{https://macho.nci.org.au/macho_photometry}} which contains millions of light curves in bands B and R. 

We selected a subset of fields 1, 101, 102, 103, and 104 containing \num{1454792} light curves for training. Similarly, we select field 10 for testing, with a total of \num{74594} light curves. MACHO observed in both bands simultaneously, therefore having two magnitudes associated with each MJD. Since we are looking to improve on Astromer 2, we maintain the single band input.  The light curves from this dataset that exhibited Gaussian noise characteristics were removed based on the criteria: $|\text{Kurtosis}| > 10$, $|\text{Skewness}| > 1$, and $\text{Std} > 0.1$. Additionally, we excluded observations with negative uncertainties (indicative of faulty measurements) or uncertainties greater than one (to maintain photometric quality). Outliers were also removed by discarding the 1st and 99th percentiles for each light curve. This additional filtering does not affect the total number of samples but reduces the number of observations when the criteria were applied.

\begin{figure}
    \centering
    \includegraphics[scale=.88]{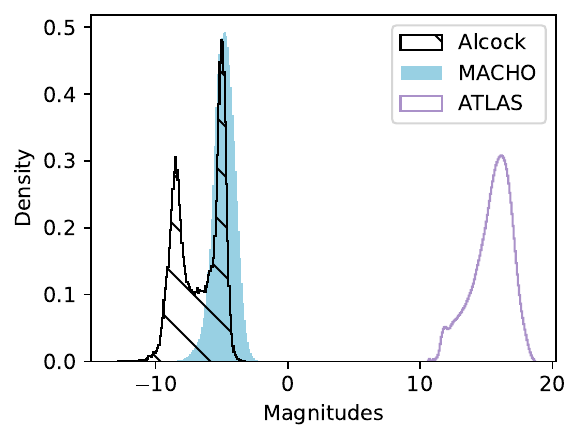}
     \caption{Magnitude distributions of the pretraining data (MACHO) and labeled sets (Alcock, ATLAS). The Alcock and MACHO distributions are similar, though Alcock is bimodal. The ATLAS data, from a different survey, shows a distinct distribution with higher variance. During training, all magnitudes are normalized, which removes the mean shifts shown here.}
    \label{fig:macho-alcock-magn}
\end{figure}

\subsection{Labeled data}
To ensure a fair comparison with Astromer 1, we used the same sample selection from the MACHO \citep[hereafter referred to as Alcock; ][]{Alcock2001Variable} and the  Asteroid Terrestrial-impact Last Alert System \citep[hereafter referred to as ATLAS; ][]{heinze2018first} labeled catalogs. The former has a similar magnitude distribution, whereas the latter differs, as shown in Fig. \ref{fig:macho-alcock-magn}.

The labeled classes in the MACHO dataset were produced through manual labeling supported by statistical analysis and feature extraction techniques, providing a curated and expert-classified dataset. For the ATLAS dataset, the labels are derived from automated classification pipelines based on template fitting and light curve features, which may introduce some uncertainty and biases inherent to automated labeling methods.

\subsubsection{Alcock}
For labeled data, we use the catalog of variable stars from \citet{Alcock2001Variable}, which contains labels for a subset of the MACHO light curves originating from 30 fields from the Large Magellanic Cloud. This labeled data will be used to train and evaluate the performance of the different embeddings on the classification task. 

The selected data comprises \num{20894} light curves, which are categorized into six classes: Cepheid variables pulsating in the fundamental (Cep\_0) and first overtone (Cep\_1), Eclipsing Binaries (EC), Long Period Variables (LPV), RR Lyrae ab and c (RRab and RRc, respectively). Table \ref{tab:alcock} summarizes the number of samples per class. We note that the catalog used is an updated version, as described in \cite{astromer}.

\begin{table}
\caption{Alcock catalog distribution.}              
\label{tab:alcock}  
\centering 
% \begin{tabular}{c c c} 
\begin{tabular}{l l r} 
\hline\hline         
Tag & Class Name & \# of sources \\ \hline
 Cep\_0 & Cepheid type I &\num{1182} \\
 Cep\_1 &Cepheid type II & \num{683} \\
 EC &Eclipsing binary & \num{6824} \\
 LPV &Long period variable &  \num{3046} \\
 RRab &RR Lyrae type ab  &  \num{7397} \\
 RRc &RR Lyrae type c &  \num{1762} \\
 Total & & \textbf{\num{20894}} \\
\hline                            
\end{tabular}
\end{table}

Figure \ref{fig:macho-alcock-magn} compares the magnitude distributions between the Alcock and MACHO datasets. The former exhibits a bimodal distribution, which aligns with the fact that it represents a subset of the light curves from MACHO fields, while the latter encompasses light curves from only five fields. 

\begin{figure}
    % \centering
    \includegraphics[scale=0.7]{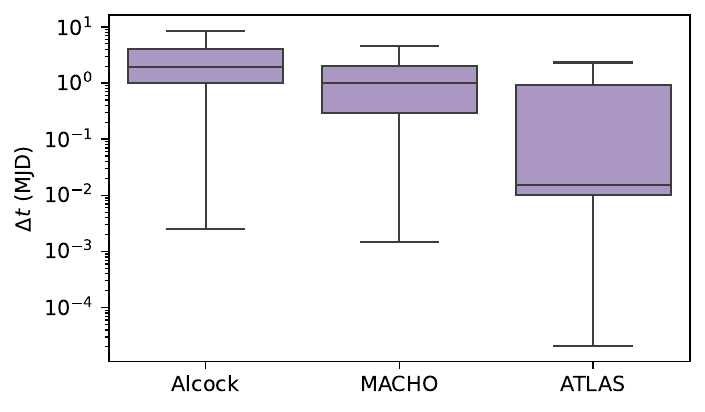}
 \caption{Observation cadence ($\Delta t$, time between consecutive points) for the MACHO, Alcock, and ATLAS datasets. The boxplots show that MACHO and Alcock have similar, regular cadences (median $\sim3-4$ days). The ATLAS dataset is distinct, with a much faster median cadence and greater variability in observation times. The y-axis uses a logarithmic scale to display the wide range of values.}
    \label{fig:macho-alcock-mjd}
\end{figure}

Similarly, we compare the distribution of time differences between consecutive observations ($\Delta t$). Figure \ref{fig:macho-alcock-mjd} shows similar distributions, with comparable ranges and means of three and four days for MACHO and Alcock, respectively.

\subsection{ATLAS}
The Asteroid Terrestrial-impact Last Alert System \citep[ATLAS; ][]{Tonry2018} is a survey developed by the University of Hawaii and funded by NASA. Operating since 2015, ATLAS has a global network telescopes, primarily focused on detecting asteroids and comets that could potentially threaten Earth. Observing in $c$ (blue), $o$ (orange), and $t$ (red) filters.

The variable star dataset used in this work was presented by \citet{heinze2018first} and includes 4.7 million candidate variable objects, included in the labeled and unclassified objects, as well as a dubious class. According to their estimates, this class is predominantly composed of $90\%$ instrumental noise and only $10\%$ genuine variable stars.

We analyze \num{141376} light curves from the ATLAS dataset, as detailed in Table \ref{tab:ATLAS}. These observations, measured in the $o$ passband, have a median cadence of $\sim$15 minutes, which is significantly shorter than the typical cadence in the MACHO dataset. This substantial difference poses a challenge for the model, as it must adapt to such a distinct temporal distribution. 

\begin{table}[h!]
\caption{ATLAS catalog distribution.}              
\label{tab:ATLAS}
\centering 
\begin{tabular}{l l r} 
\hline\hline         
Tag & Class Name & \# of sources \\
\hline
CB & Close Binaries &  \num{80218} \\
DB & Detached Binary &  \num{28767} \\
Mira & Mira &  \num{7370} \\
Pulse &RR Lyrae, $\delta$-Scuti, Cepheids &  \num{25021} \\
Total & & \textbf{\num{141376}}\\
\hline                            
\end{tabular}
\end{table}

As done in \citet{astromer} and to standardize the labels with other datasets, we combine detached eclipsing binaries identified by full or half periods into the close binaries (CB) category and similarly merge detached binaries (DB). However, objects with labels derived from Fourier analysis are excluded, as these classifications do not directly align with astrophysical categories.

\subsection{MACHO vs ATLAS}\label{sec:machovsatlas}
Figures \ref{fig:macho-alcock-magn} and \ref{fig:macho-alcock-mjd} illustrate the distributional differences between the unlabeled MACHO dataset and the labeled subsets discussed earlier. While the magnitudes show a notable shift between MACHO and ATLAS, our training strategy normalizes the light curves to a zero mean. As a result, the relationships between observations take precedence over the raw magnitude values. Consequently, we do not expect a substantial performance drop when transitioning between datasets. However, for $\Delta t$, the smaller values of $\Delta t$ present a significant challenge, as the model must extrapolate and account for fast variations to capture short-time information effectively. We evidence this in our first results from Astromer 2, where the F1 score on the ATLAS dataset was lower compared to MACHO when having fewer labels for classification. 

\section{Results}\label{sec:results}
The pretraining task of reconstructing the probed magnitudes is an essential component that allows the model to learn meaningful representations. Reconstructing probed magnitudes can be evaluated as a downstream regression task on labeled datasets, where the model's ability to predict the magnitudes can be assessed. Here we evaluate the potential of the representation in terms of regression.

\begin{figure}
    \centering
    \includegraphics[scale=0.8]{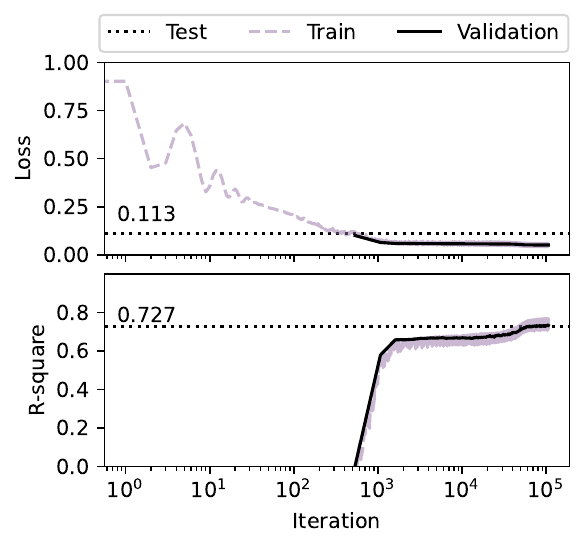}
     \caption{Pretraining learning curves for Astromer 2, showing reconstruction loss (RMSE) versus training epochs. Both training (solid blue) and validation (orange) losses decrease and converge, indicating stable training. The final performance on the held-out test set is marked by the dotted line.}
    \label{fig:trastromerv1}
\end{figure}

Figure \ref{fig:trastromerv1} shows the learning curves from the pretraining of Astromer 2. The training took approximately 3 days using 4 A5000 GPUs. The model achieved an $R^2$ of $0.73$\footnote{$R^2$ (coefficient of determination) measures the proportion of variance in the observed data explained by the model. It ranges from ($-\infty$) to 1, with 1 indicating perfect prediction.}with a root mean squared error (RMSE) of 0.113 on the probed subset. Astromer 1 had an RMSE of 0.148, making Astromer 2 0.035 better in terms of reconstruction error.  

\subsection{Downstream setup}
Similar to Astromer 1, this work evaluates the embeddings across various scenarios, while controlling the number of samples per class (SPC). When there are few SPC, the quality of the embeddings becomes crucial, as they must capture fundamental features that enable the model to make predictions based solely on the shape of the light curves, without relying on additional information.

Using the labeled datasets, we construct three training scenarios with limited data by randomly sampling 20, 100, and 500 SPC to assess performance on downstream tasks. When the number of available labels is insufficient, sampling is performed with replacement. To account for variability in our experiments, we generate three folds for each scenario. To ensure fairness across scenarios, we evaluate all models on a shared 3-fold test dataset consisting of 1000 SPC per fold.
Hence, the models trained on 20, 100, and 500 SPC were evaluated against this common dataset.

In the initial stage of the downstream pipeline, we performed finetuning. This involves loading the pretrained weights and adapting the encoder parameters to the target domain. Finetuning followed the same pretraining setup, predicting a random probed subset containing $50\%$ of the magnitudes per sequence.

\subsection{Finetuning}
\begin{figure}
    \centering
    \includegraphics[scale=0.6]{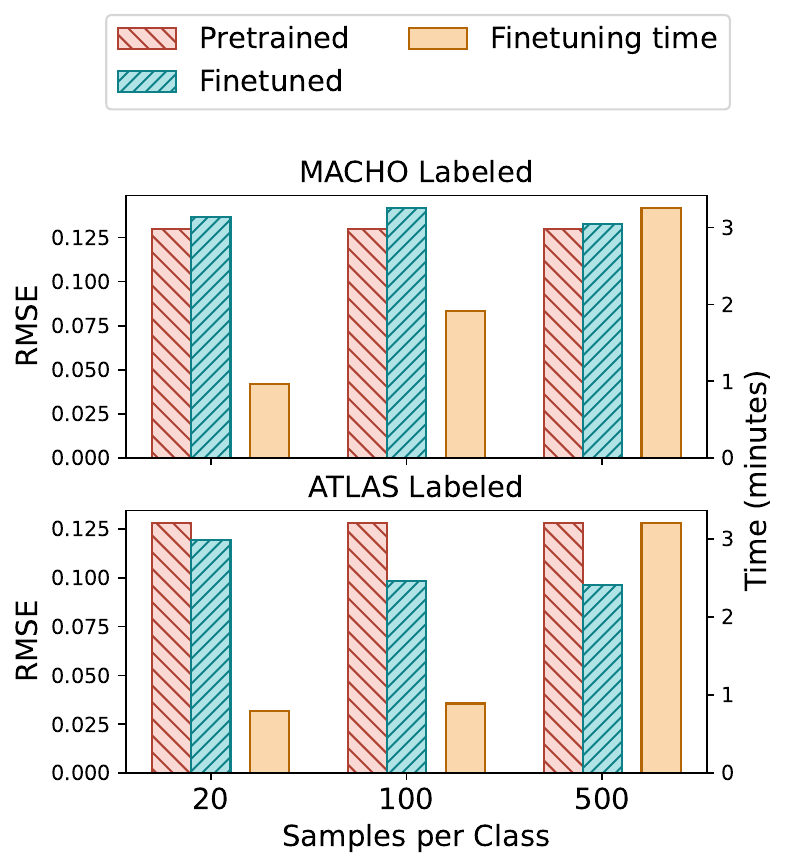}
    \caption{Finetuning performance and computational cost. The dashed bars show reconstruction error (RMSE, left y-axis) and the solid bars show elapsed time (minutes, right y-axis) for finetuning on the Alcock and ATLAS datasets with varying samples per class (SPC). Finetuning significantly reduces error for ATLAS, while the time required remains low across all scenarios.}
    \label{fig:ft_metrics}
\end{figure}

Figure \ref{fig:ft_metrics} presents the RMSE results for each scenario. The reported values are calculated across the total number of observations without masking, allowing for a fair comparison as the error could be biased by the random masking selection. 
As shown in Fig. \ref{fig:ft_metrics}, finetuning the model on the Alcock dataset does not result in significant improvements, indicating that the pretrained model already captures most of the relevant information, despite the out-of-distribution modality discussed in Sect. \ref{sec:machovsatlas}. In contrast, finetuning on ATLAS leads to a notable improvement. Specifically, with 100 SPC, we observe a $23\%$ reduction in RMSE compared to the pretrained model. However, the performance improvement between 100 and 500 SPC is minimal, with only small variations.

The most computationally intensive scenario takes approximately three minutes to finetune, which is significantly faster than the days required for pretraining. While the time for MACHO increases with more samples, in ATLAS, we observe almost no variation between 20 and 100 SPC. This is because MACHO samples are longer than ATLAS samples, resulting in a more substantial increase in the number of windows as the number of SPC grows. This explains the more significant rise in computing time when going from 20 to 100 MACHO SPC.

\subsection{Visualizing reconstruction}
\begin{figure}
    \centering
    \includegraphics[scale=0.7]{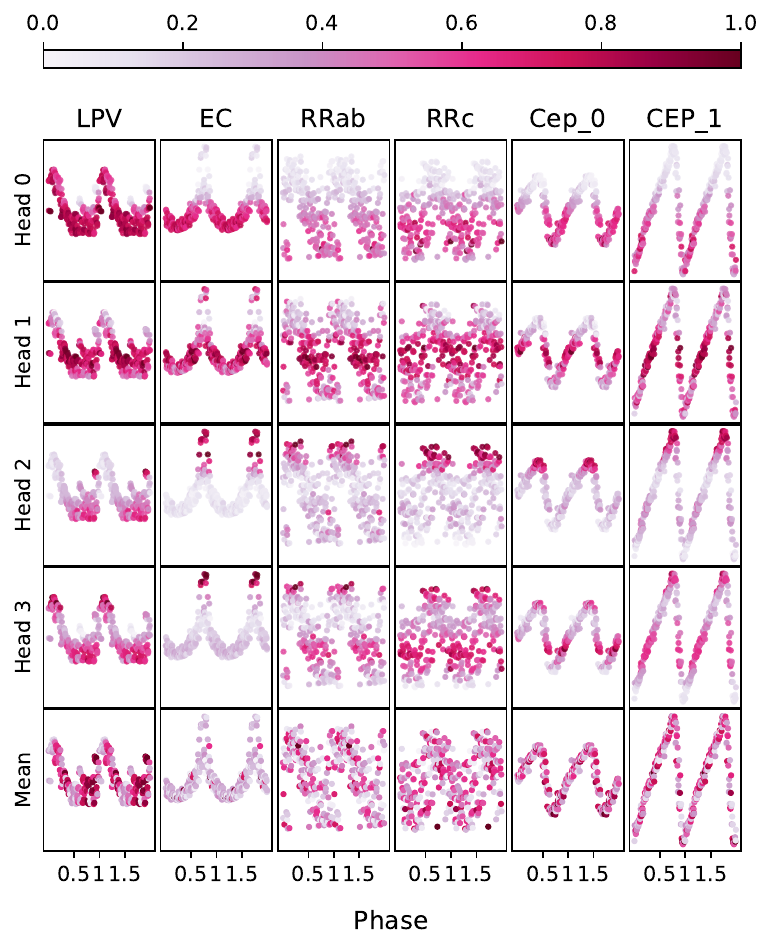}
    \caption{Visualization of attention weights from the first encoder block on sample light curves from the Alcock dataset. Each of the first 8 plots shows the average attention from one head (colored bar, top), with the final plot showing the mean across all heads. The light curves are folded by their period for clarity. The model consistently focuses attention on points of maximum and minimum brightness.}
    \label{fig:attention_vis}    
\end{figure}

\begin{figure}
    \centering
    \includegraphics[scale=0.7]{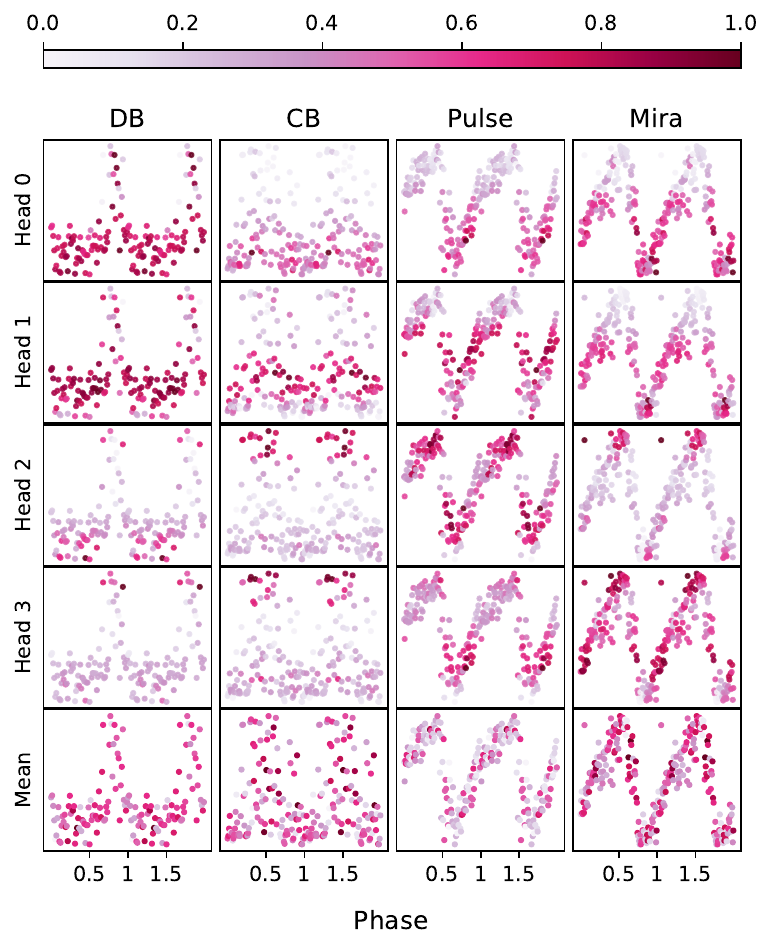}
    \caption{Same as Fig. \ref{fig:attention_vis}, but for sample light curves from the ATLAS dataset. The model again learns to focus on the extrema of the light curves, demonstrating a consistent feature-learning strategy across different datasets.}
    \label{fig:attention_vis_altas}    
\end{figure}

% To gain insight into Astromer's representations, we visualize the attention values after finetuning. Figure \ref{fig:attention_vis} presents two examples showing the mean attention weights from each attention head, along with the mean across all attention heads. 
Figures \ref{fig:attention_vis} and \ref{fig:attention_vis_altas} show the mean attention weights from each attention head, along with the mean across all heads, for sample light curves from the Alcock and ATLAS datasets, respectively. 
For each dataset, we present one representative example for each class to illustrate the model's focus. For visualization purposes, the light curves were folded by their period; however, Astromer does not receive folded inputs during processing. We display the attention from the first attention block, as its interpretation is more intuitive compared to intermediate layers, where attention is computed over more abstract embeddings. As shown in both figures, each attention head learns to focus on different parts of the sequence. Notably, there is a consistent pattern where attention is most strongly concentrated at points of maximum and minimum brightness, suggesting these are identified as key features for reconstruction and characterization. This demonstrates a consistent feature-learning strategy across different datasets and variable star types.
% For visualization purposes, light curves associated with the average between heads were folded; however, Astromer does not receive folded inputs during processing. We display only the first attention block, as it is more intuitive. This contrasts with intermediate layers, where attention is computed over abstract embeddings.
% Figures \ref{fig:attention_vis} and \ref{fig:attention_vis_altas} show that each attention head focuses on different parts of the sequence. In particular, attention appears to focus most strongly at maximum and minimum brightness points suggesting these are key features for reconstruction. 

% \subsection{Forecasting}

\subsection{Classification}
Evaluating classification performance is crucial for assessing the overall effectiveness of Astromer, as it serves as a common benchmark for evaluating the quality of embeddings. After fine-tuning on labeled subsets of 20, 100, and 500 SPC, the encoder is frozen, meaning its weights are no longer updated

Astromer is used to extract the representation, which is fed to another classifier model. The same labeled data is used to train a classifier. In this setup, only the classifier section receives label-based gradients, while the encoder focuses exclusively on capturing dependencies between observationns.

\begin{figure}
    \centering
    \includegraphics[scale=0.4]{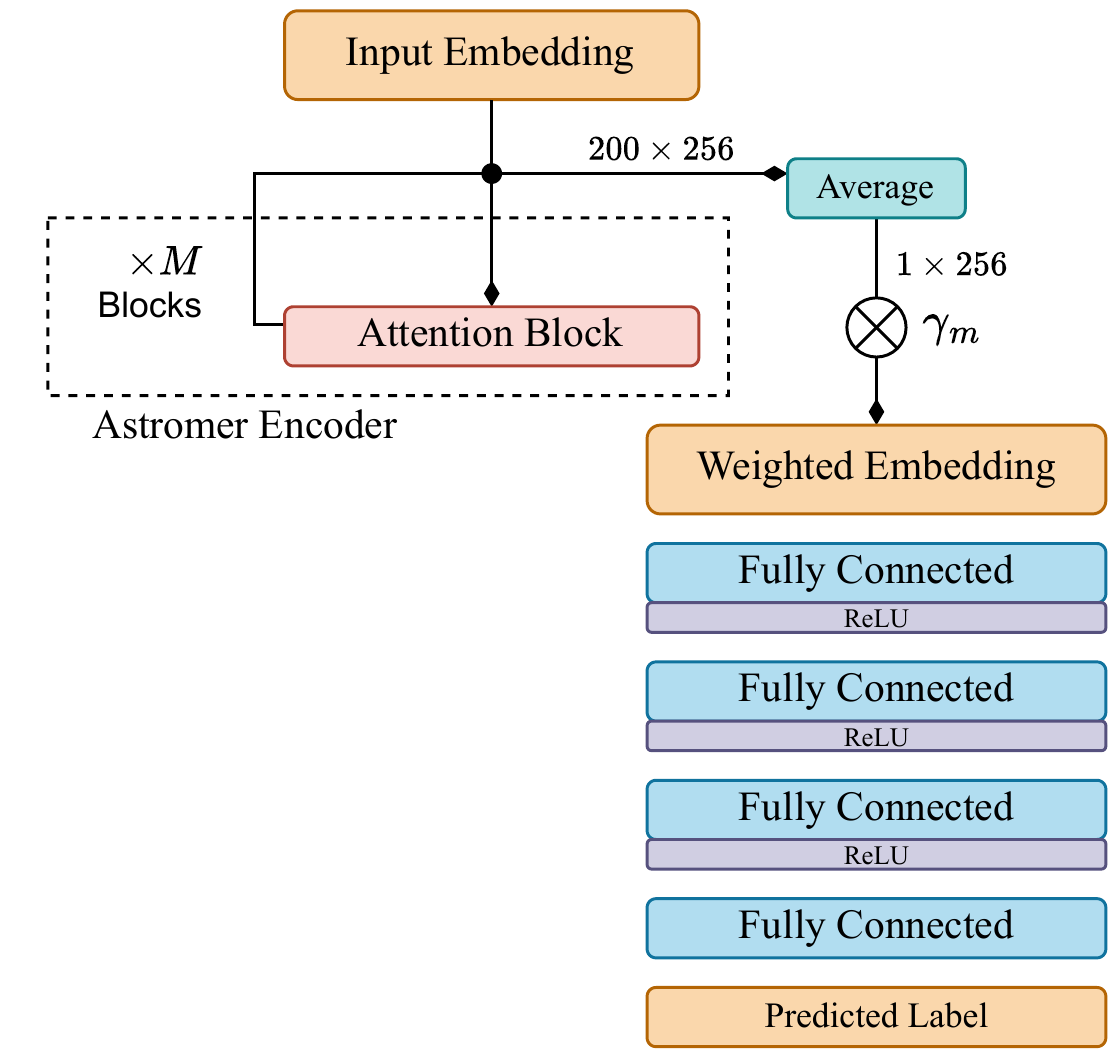}
    \caption{The downstream classification head architecture. A weighted-average embedding from the Astromer encoder is passed through a feed-forward network (FFN) with three hidden layers to produce the final class prediction. Only the weights of this classifier head are trained on the labeled data.}
    \label{fig:clf-arch}
\end{figure}

Per-sample embeddings were generated by averaging the attention vectors from the encoder, with trainable parameters $\gamma_0,...,\gamma_m$ weighting the outputs of each block. The resulting embedding is then passed through a feed-forward network consisting of three hidden layers with 1024, 512, and 256 units, respectively, each using ReLU activation. A fully connected layer without activation predicts the final label, as shown in Figure \ref{fig:clf-arch}.

\begin{figure}
    \centering
    \includegraphics[scale=0.65]{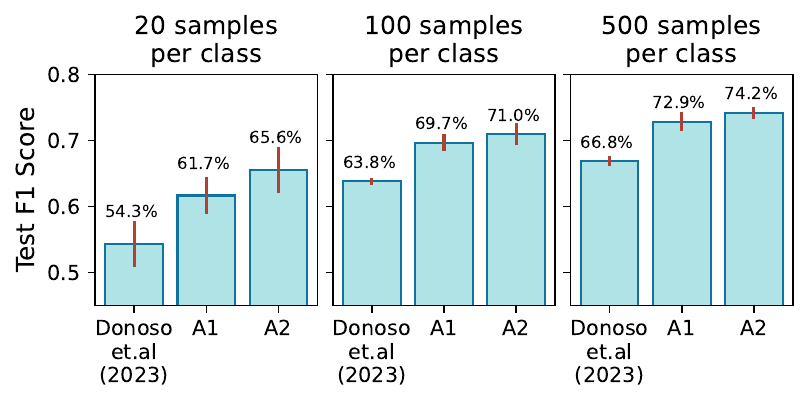}
     \caption{Classification performance (F1-score) on the Alcock (MACHO) dataset. The plot compares Astromer 2 (A2) against Astromer 1 (A1) and the published results from \citet{astromer} across scenarios with 20, 100, and 500 labeled samples per class (SPC). Astromer 2 consistently outperforms previous versions, especially in low-data regimes.}
    \label{fig:macho-clf}
\end{figure}

Figure \ref{fig:macho-clf} presents the F1 scores for the Alcock dataset across different scenarios. We focus our analysis on the improvements over our previous model, rather than re-evaluating against classical models, as this benchmark was already established in our initial work \citep{astromer}. For a direct comparison, the figure includes the F1 scores from Astromer 1, using both weighted per-sample embeddings (A1) and non-weighted embeddings (A2) as detailed in our previous paper.
% Figure \ref{fig:macho-clf} presents the F1 scores for the Alcock dataset across different scenarios. For comparison, it also includes F1 scores from Astromer 1, evaluated on the same dataset, using both weighted per-sample embeddings (v0) and non-weighted embeddings as detailed in \citet{astromer}.

\begin{figure}
    \centering
    \includegraphics[scale=0.65]{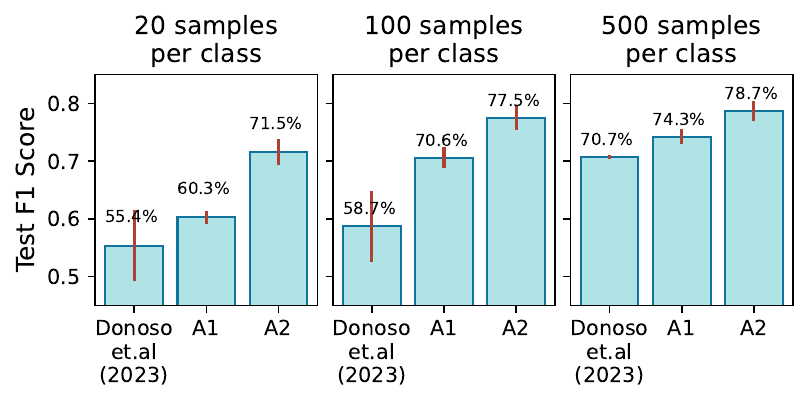}
    \caption{Same as Fig. \ref{fig:macho-clf}, but for the ATLAS dataset. The performance improvement of Astromer 2 is even more pronounced on this out-of-distribution dataset, where it significantly surpasses all previous results, even when trained with only 20 SPC.}
    \label{fig:atlas-clf}
\end{figure}

The improvements are most pronounced when evaluating classification performance on the ATLAS dataset. As depicted in Figure \ref{fig:atlas-clf}, the new version of Astromer exhibits a significant advantage in generalizing to other datasets. With just 20 SPC, Astromer achieves a F1-score improvement of over $15\%$. Thus, Astromer's performance with 20 SPC surpasses the results previously reported 500 SPC.

Weighted per-sample embeddings play a critical role by allowing the model to use intermediate representations instead of depending exclusively on the final one. During pretraining, the encoder focuses on reconstructing magnitudes from the last embedding, which could result in representations tailored to the reconstruction task rather than optimized for discrimination.

\begin{figure}
    \centering
    \includegraphics[scale=0.8]{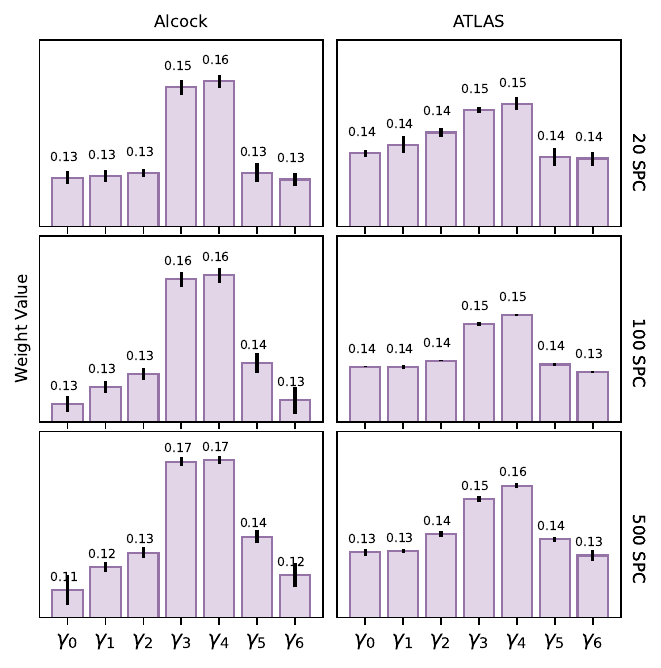}
 \caption{Learned gamma ($\gamma$) weights for the weighted-average embedding. Each $\gamma_i$ scales the contribution of the output from block i of the encoder ($\gamma_0$ for the input). The classifier learns to assign higher importance to intermediate layers (2-5) rather than the initial or final layers, especially as more training data becomes available.}
    \label{fig:gammaweights}
\end{figure}

To examine the role of intermediate embeddings, we plot the gamma parameters after training the classifier on both the Alcock and ATLAS datasets. As shown in Fig. \ref{fig:gammaweights}, the weights assigned to intermediate embeddings are higher than those for the initial or final ones. This disparity becomes more pronounced as the number of training samples increases.

\begin{figure}
    \centering
    \includegraphics[scale=0.7]{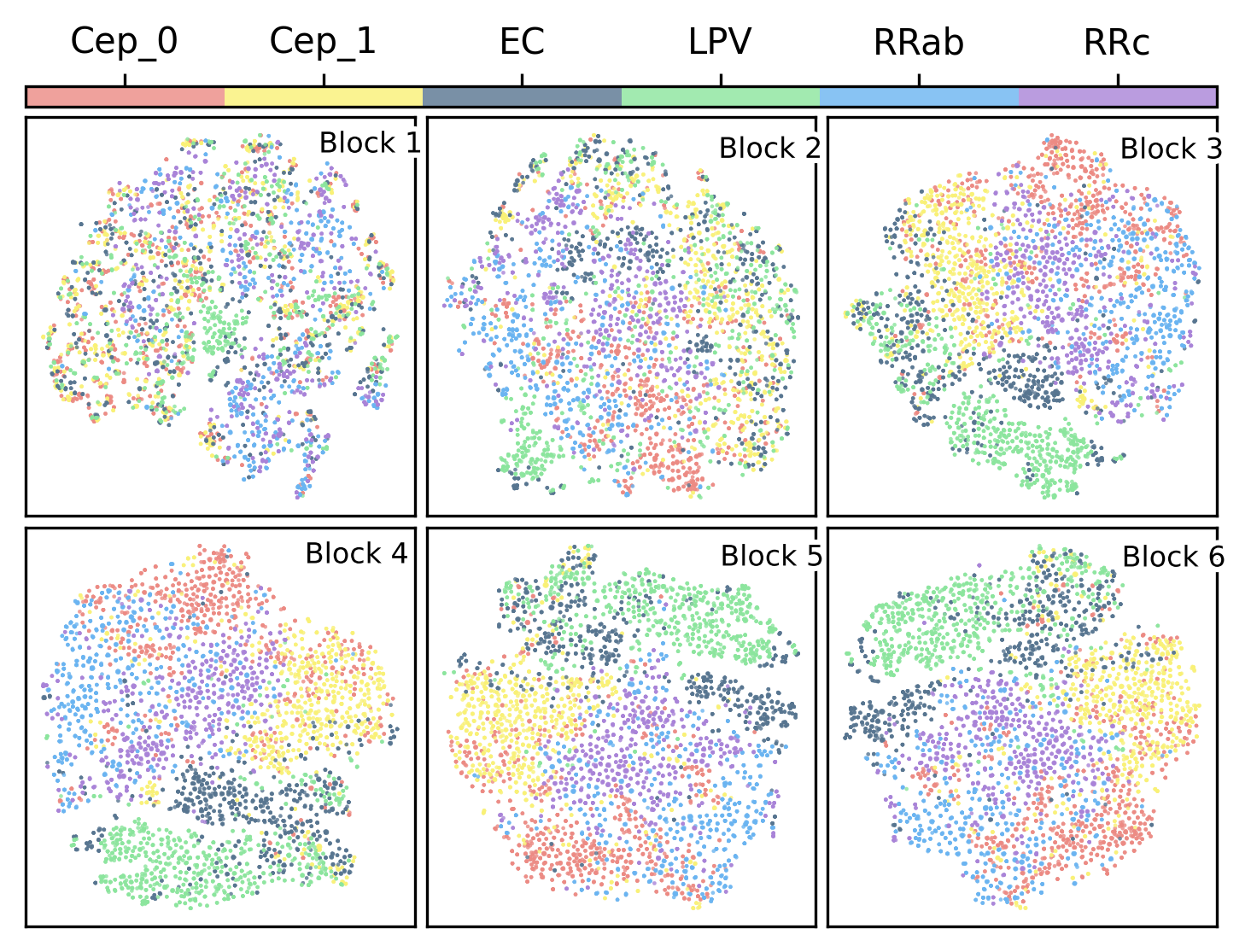}
    \caption{t-SNE visualization of embeddings from each of the six encoder blocks for the Alcock (MACHO) test set. Each point is a light curve, colored by its true class. The plots show that class structure emerges and improves in the intermediate and deeper layers, even though the model was pretrained without any label information.}
    \label{fig:alcock-emb}
\end{figure}

\begin{figure}
    \centering
    \includegraphics[scale=0.7]{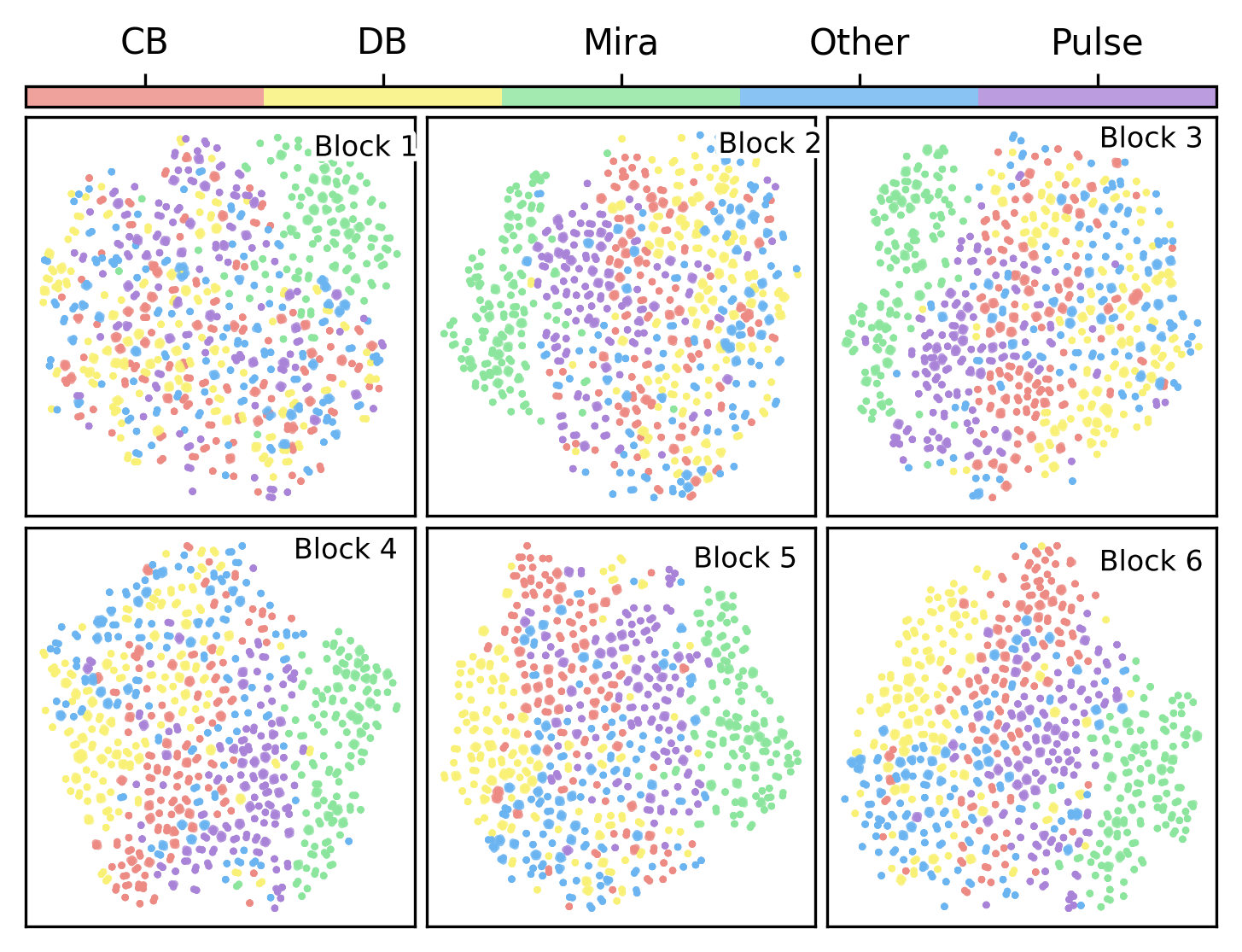}
    \caption{Same as Fig. \ref{fig:alcock-emb}, but for the ATLAS test set embeddings. A similar trend is observed where embeddings from deeper layers provide better class separation, demonstrating the model's ability to learn meaningful structures for different data distributions.}
    \label{fig:atlas-emb}
\end{figure}

Figures \ref{fig:alcock-emb} and \ref{fig:atlas-emb} present t-SNE projections of the test set embeddings, derived from the output of each attention block for the first data fold. Unlike standard text-based BERT models, our architecture does not employ a [CLS] token to obtain a summary representation of the sequence. Therefore, an aggregation step is necessary to produce a single vector for each time series. To achieve this, we reduce the (200, 256) output tensor from each block by averaging across the 200 time steps, resulting in a representative (1, 256) vector.

This averaging strategy was selected over other alternatives, such as concatenation and the maximum, as it consistently produces more coherent clusters for visualization. A detailed quantitative justification supporting this decision is provided in the Appendix \ref{visualization}

The first thing to notice is that there is evidence that Astromer properly separates classes in both the Alcock and ATLAS datasets. Recall that these embeddings were trained solely on light curve reconstruction, without any information about the labels.

A key advantage of our self-supervised approach is its ability to enable object retrieval and similarity searches directly in the embedding space without requiring labeled data. As shown in Figures \ref{fig:alcock-emb} and \ref{fig:atlas-emb}, the model organizes light curves with similar characteristics close to one another, which facilitates effective clustering and retrieval. This capability is particularly valuable in astronomical datasets where labeling is costly and incomplete, providing a powerful tool for exploratory data analysis and discovery beyond classification tasks.

Classes are separated in different ways depending on the block. For both Alcock and ATLAS, the first block does not seem to separate classes at all, while other blocks exhibit better discrimination. This aligns with the $\gamma$ parameters we introduced to weight each block's output during classifier training (see Figure \ref{fig:gammaweights}).

\begin{figure}
    \centering
    \includegraphics[scale=0.58]{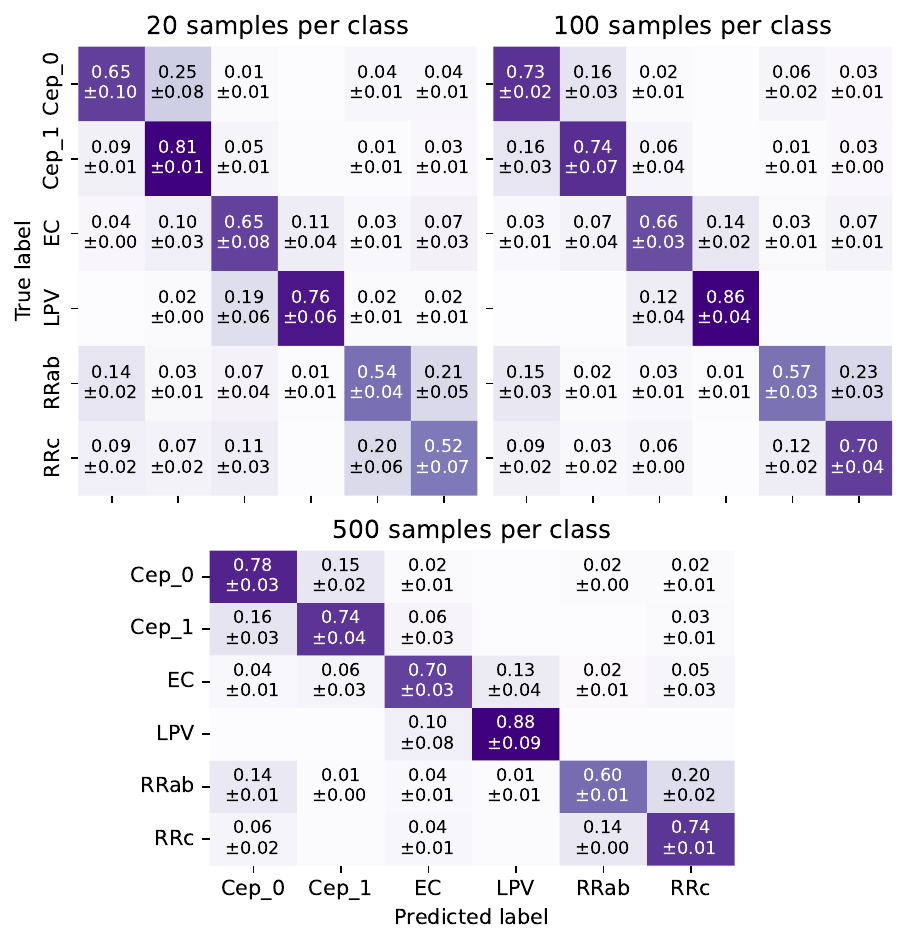}
     \caption{Confusion matrices for classification on the Alcock test set, for models trained with 20, 100, and 500 samples per class (SPC). As more labeled data is used, the diagonal (correct classifications) becomes stronger. The most persistent confusion occurs between the RR Lyrae subtypes (RRab and RRc).}
    \label{fig:cm_alcock}
\end{figure}

\begin{figure}
    \centering
    \includegraphics[scale=0.58]{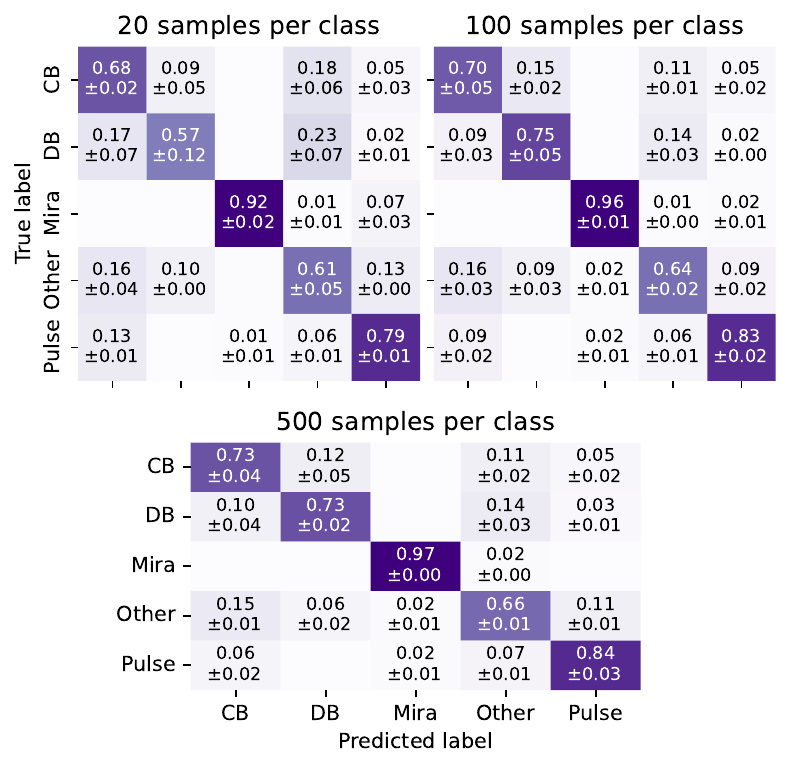}
    \caption{Same as Fig. \ref{fig:cm_alcock}, but for the ATLAS test set. Performance improves with more SPC, but significant confusion remains involving the 'Other' class, which is inherently diverse and overlaps with defined classes in the feature space.}
    \label{fig:cm_atlas}
\end{figure}

Alternatively, we can observe the effect of class separation in the confusion matrices in Figures \ref{fig:cm_alcock} and \ref{fig:cm_atlas}. Specifically, for the Alcock dataset, the main confusion occurs between RR Lyrae types ab and c. A similar pattern is observed in the projection of Figure \ref{fig:alcock-emb}, where the blue (RRab) and purple (RRc) points are mixed together. In ATLAS, there is an \textit{Other} class, which can be particularly confusing for the model. As observed in Figure \ref{fig:atlas-emb}, the blue points corresponding to the \textit{Other} class are sparsely distributed in the embedding space.

\section{Conclusion}\label{sec:conclusion}
This paper presents the updated version of Astromer, a self-supervised model designed to extract general-purpose embeddings from light curve data. We demonstrate that Astromer 2 outperforms its predecessor, Astromer 1, across multiple scenarios, including both the Alcock and ATLAS datasets. The key improvements in classification performance, especially when trained with limited labeled samples, highlight the informative power of the embeddings in discriminating between different classes.

The results show that Astromer 2 achieves significant gains in F1 score, especially with small training sets, underscoring the effectiveness of the weighted per-sample embeddings and the model’s ability to generalize across different datasets. The analysis of attention weights further reveals that intermediate embeddings contribute meaningfully to the model's performance, focusing more on certain parts of the input data during the classification task.

These findings confirm the potential of Astromer as a robust tool for light curve analysis, showing that self-supervised learning can provide valuable insights into astronomical data. Future directions may explore the integration of multi-modal data and more complex attention mechanisms, though the simplicity and efficiency of the current approach remain significant advantages. Future work on Astromer will focus on incorporating multiband data, either as an additional feature embedding or by directly constraining the embedding space. Additionally, we plan to train Astromer on the entire survey, bringing in more data during pretraining and potentially capturing more informative representations. 

\section{Ethical and Practical Considerations}
Large-scale models require significant computational resources, highlighting the need for optimization. According to \cite{lannelongue2021green}, training Astromer 2 resulted in $32.29$ kg of $\text{CO}_2$ emissions, which is equivalent to a 195.96 km trip in a passenger car.

To mitigate this environmental impact, we provide pre-trained weights in our repository, enabling users to build upon our model without the need for full retraining. If you have new data, we encourage you to share your pre-trained models as a pull request in our repository.

Code, weights, and data can be found in our official organization: \url{https://github.com/astromer-science}. Additionally, we provide a website with user practical information \url{https://www.stellardnn.org/projects/astromer/index.html}.

% \begin{acknowledgements}
% This work ....      
% \end{acknowledgements}
\bibliographystyle{aa} % style aa.bst
\bibliography{references} % your references Yourfile.bib

\appendix

\section{Embedding Visualization Method}\label{visualization}
To select an optimal method for visualizing the class-based clustering of our pre-trained encoder embeddings, we conducted a quantitative comparison of three distinct aggregation strategies. The methods evaluated were: averaging the token embeddings across the time dimension, concatenating all token embeddings, and applying max pooling across the time dimension.

We used the Silhouette Score with a cosine metric as the evaluation criterion to determine which method produced the most coherent and well-separated clusters. The score for each method was calculated at the output of each of the six attention blocks in the encoder.

\begin{figure}[h]
    \centering
    \includegraphics[scale=.7]{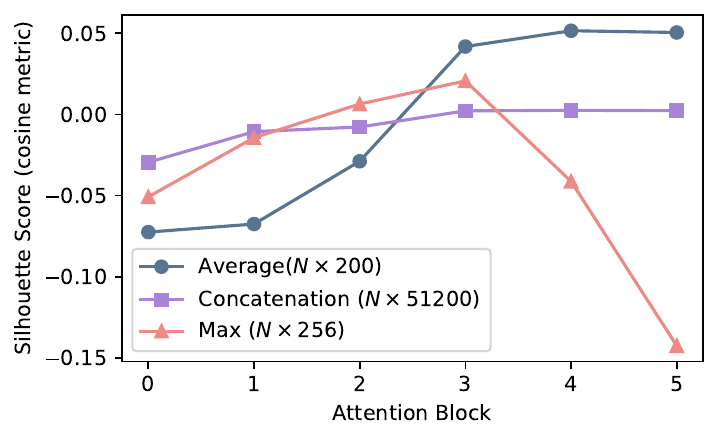}
    \caption{Silhouette Score, calculated with a cosine metric, as a function of the attention block number for three embedding aggregation methods. The methods compared are averaging (blue circles) , concatenation (purple squares) , and max pooling (red triangles). The averaging method consistently achieves the highest score in the final layers, indicating it is the most suitable strategy for producing coherent visualizations of class clusters.}
    \label{fig:silhuette}
\end{figure}

The results of this analysis are presented in Figure \ref{fig:silhuette}. The key findings from this comparison are:

\begin{itemize}
    \item The averaging method consistently achieves the highest Silhouette Score in the later, more specialized layers of the encoder (blocks 3, 4, and 5) , peaking at a score above $0.05$. This indicates that averaging produces the most distinct visual clusters.
    \item The max pooling method shows competitive performance in the middle layers but degrades sharply in the final blocks, suggesting it does not capture the final, refined embedding structure as effectively.
    \item The concatenation method results in a score of approximately zero across all blocks, implying it does not produce a useful clustering structure for visualization.
\end{itemize}

Based on this evidence, we selected the averaging strategy for all embedding visualizations presented in this work.

\section{Examples of Masked Value Reconstruction}
To provide a qualitative assessment of the model's ability to learn the underlying structure of the light curves, we present examples of its reconstruction performance on masked input data. 

The following figures show a detailed analysis for a selection of variable star classes: Cepheid (Cep\_0 , Cep\_1 ), RR Lyrae (RRab , RRc ), Eclipsing Binary (EC ), and Long Period Variable (LPV ). Each figure displays a grid corresponding to three randomly selected light curves from its respective class.

Each row in the grid provides three different views of the model's reconstruction for a single sample:
\begin{enumerate}
    \item Left Column (Time Series Reconstruction): This panel displays the original light curve in the time domain. It shows the unmasked observations (Original Visible ), the true values of the masked points (Original Masked ), and the model’s corresponding predictions (Model Prediction ).
    \item Center Column (Prediction vs. True): This panel offers a direct one-to-one comparison of the predicted versus true magnitudes for the masked data points. A dashed line represents a perfect prediction, and the Root Mean Square Error (RMSE) is reported for each sample.
    \item Right Column (Folded Reconstruction): This panel presents the same reconstruction of masked points but in phase space, with the light curve folded by its dominant period. This view is crucial for assessing how well the model captures the periodic nature of the stellar variability.
\end{enumerate}

\begin{figure*}
    \centering
    \includegraphics[scale=.7]{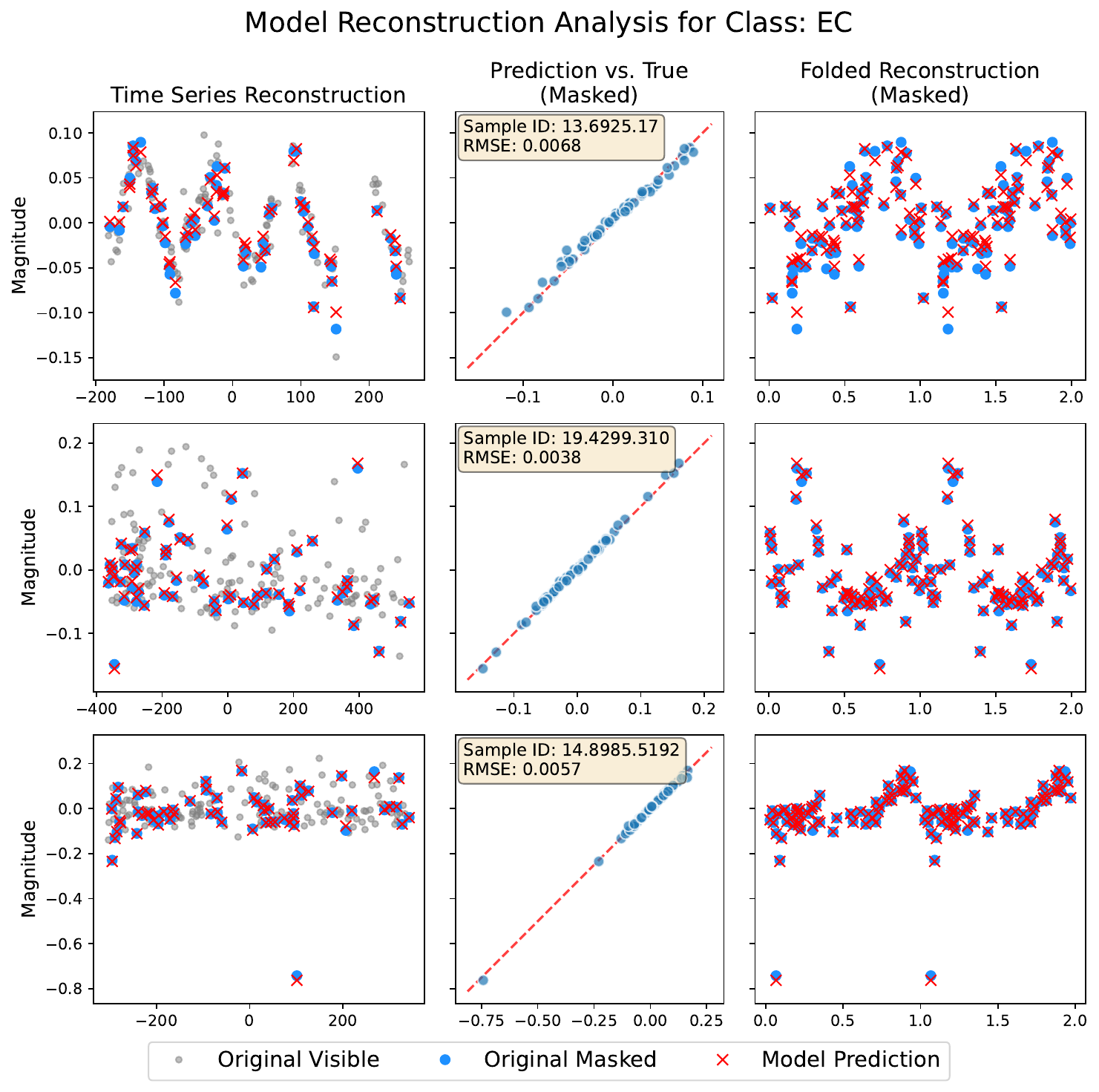}
\end{figure*}
\begin{figure*}
    \centering
    \includegraphics[scale=.7]{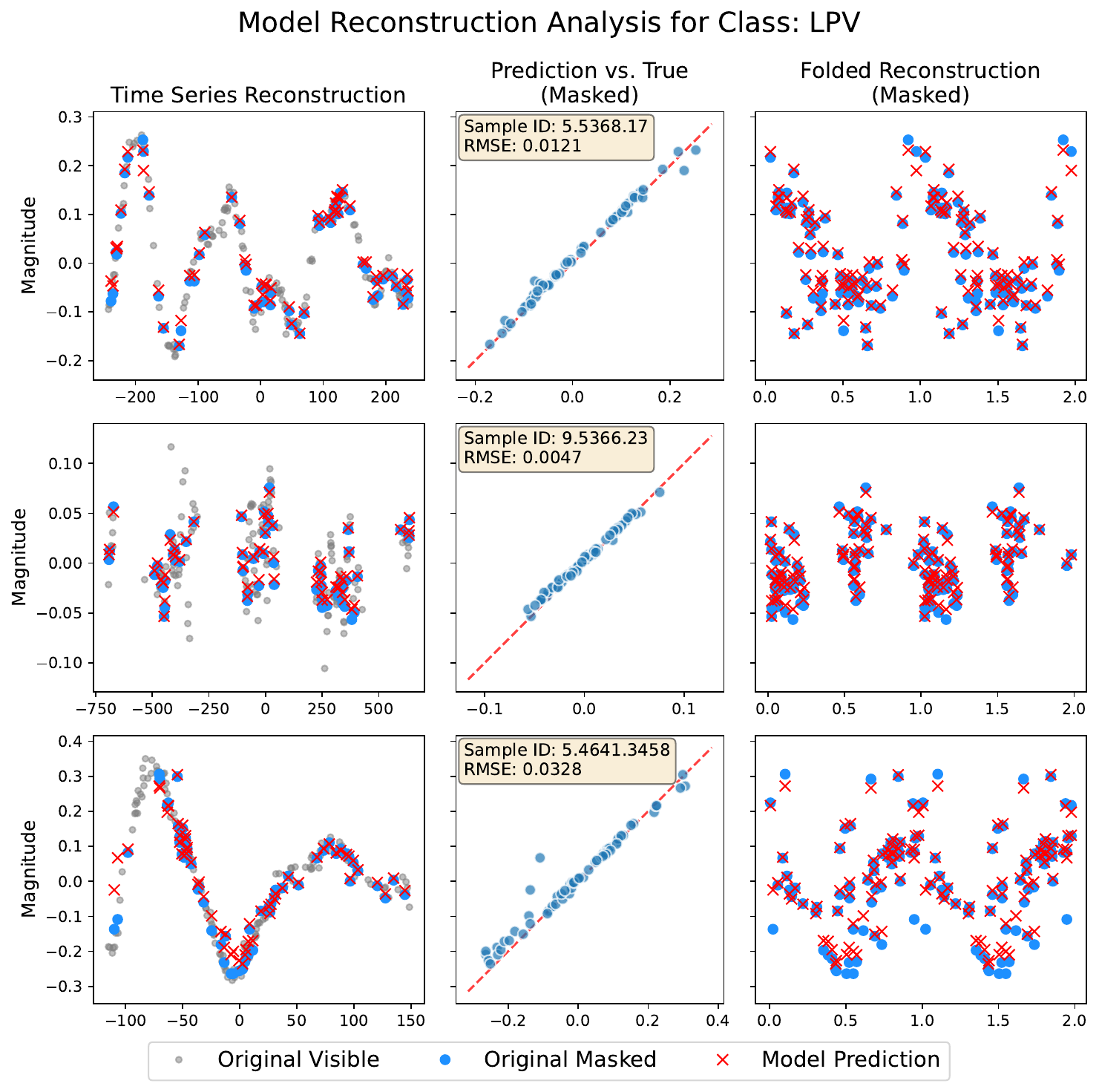}
\end{figure*}
\begin{figure*}
    \centering
    \includegraphics[scale=.7]{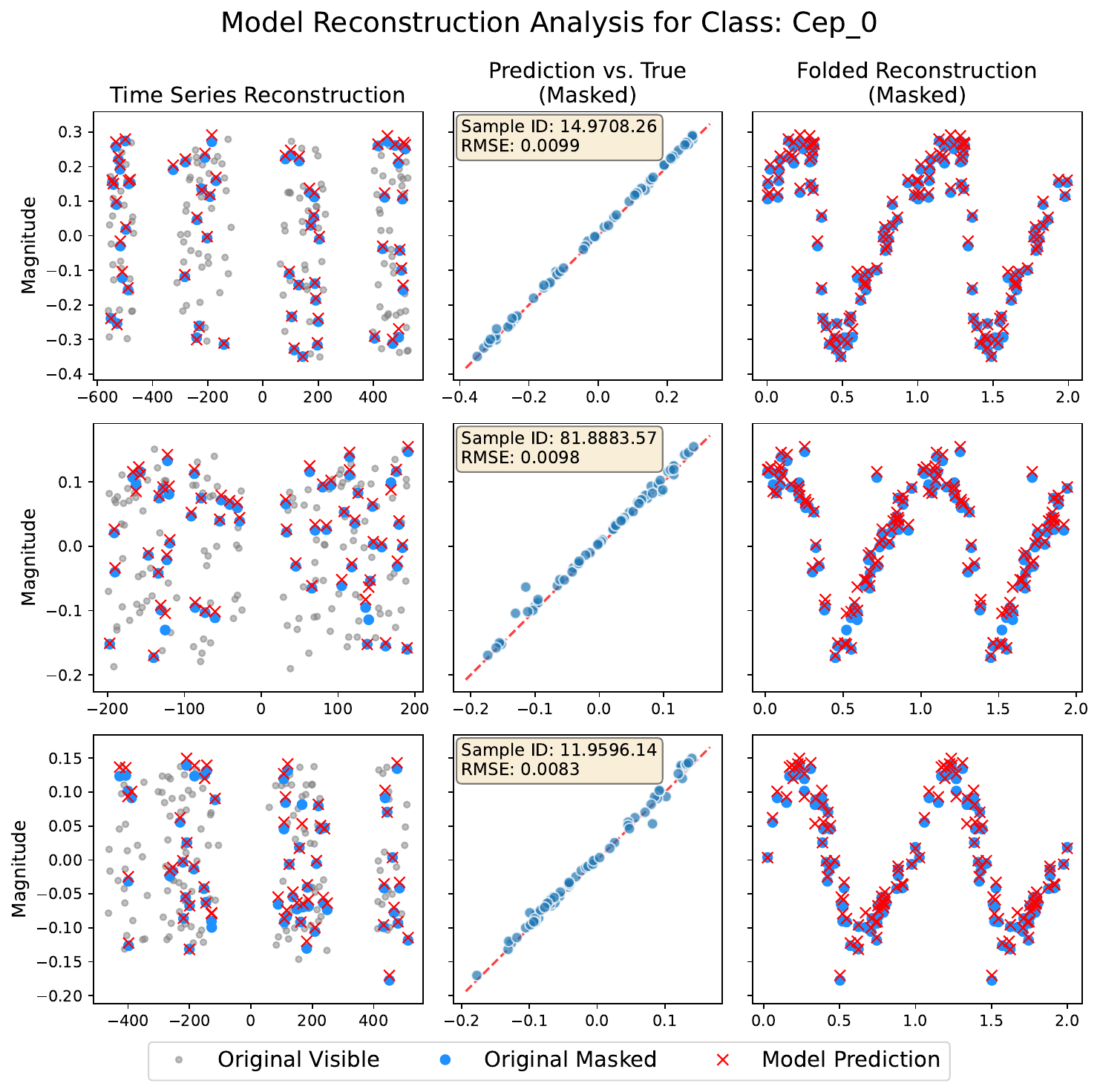}
\end{figure*}
\begin{figure*}
    \centering
    \includegraphics[scale=.7]{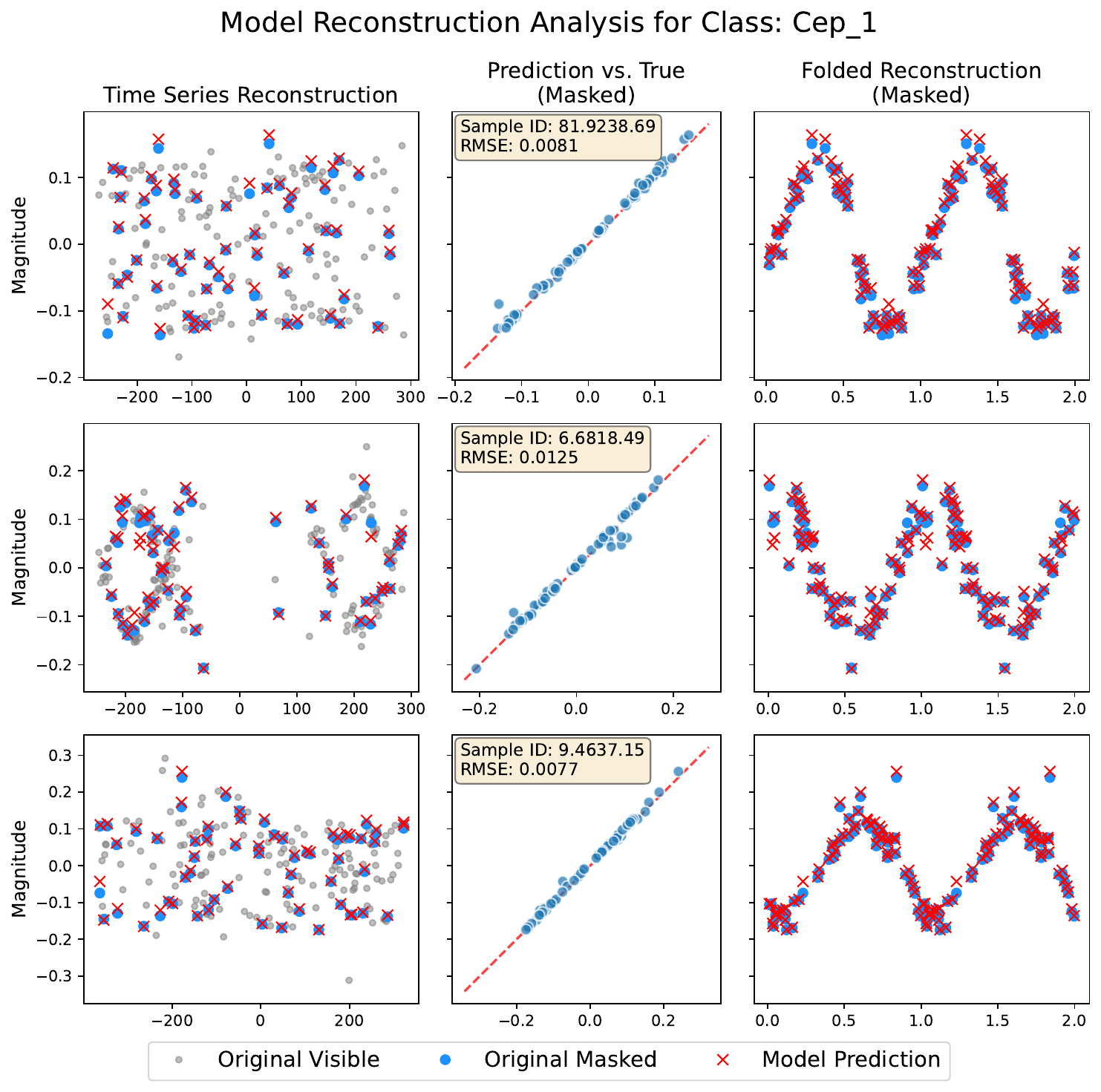}
\end{figure*}
\begin{figure*}
    \centering
    \includegraphics[scale=.7]{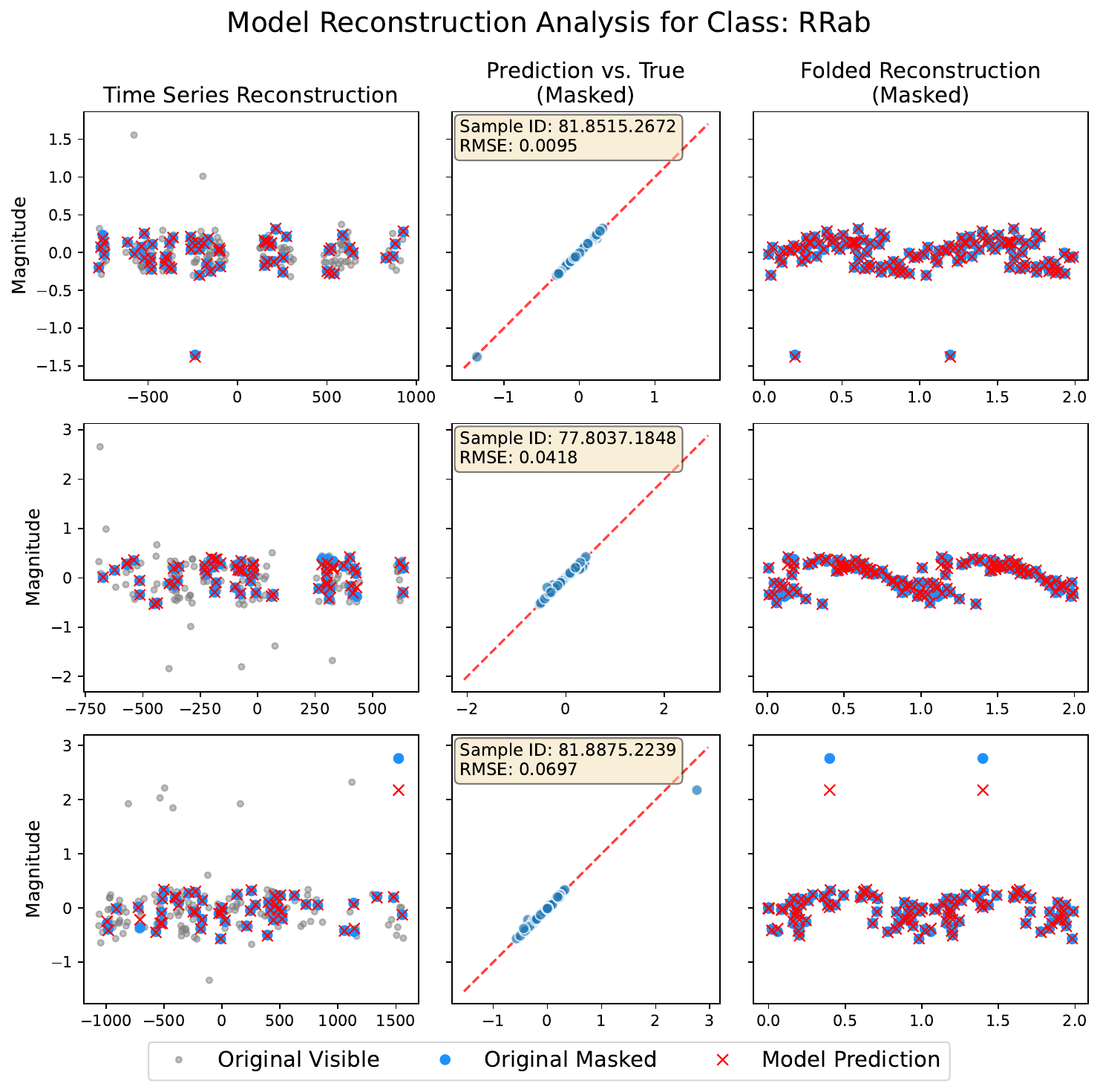}
\end{figure*}
\begin{figure*}
    \centering
    \includegraphics[scale=.7]{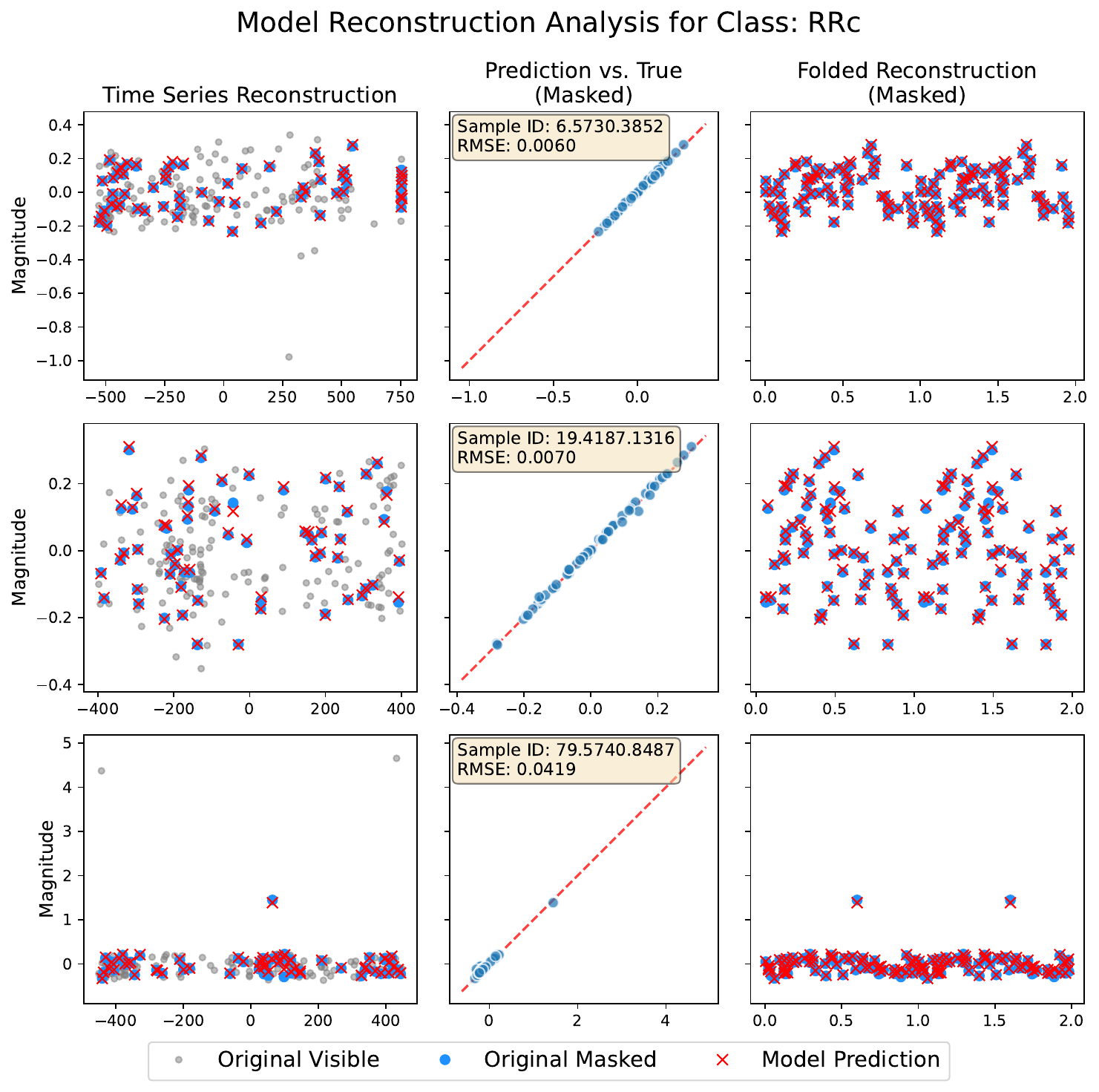}
\end{figure*}

\end{document}